\documentclass[11pt,journal,draftcls,onecolumn,peerreviewca]{IEEEtran}
\usepackage{amsmath,amsfonts}
\usepackage{algorithmic}
\usepackage{algorithm}
\usepackage{array}
\usepackage[caption=false,font=normalsize,labelfont=sf,textfont=sf]{subfig}
\usepackage{textcomp}
\usepackage{stfloats}
\usepackage{url}
\usepackage{verbatim}
\usepackage{graphicx}
\usepackage{cite}
\usepackage{amssymb}
\usepackage[numbers,sort&compress]{natbib}
\begin{document}

\title{DM-SBL: Channel Estimation under Structured Interference}
\author{Yifan Wang, Chengjie Yu, Jiang~Zhu, Fangyong Wang, Xingbin Tu, Yan Wei and Fengzhong Qu}

\markboth{Journal of \LaTeX\ Class Files,~Vol.~14, No.~8, August~2021}%
{Shell \MakeLowercase{\textit{et al.}}: A Sample Article Using IEEEtran.cls for IEEE Journals}


\maketitle

\begin{abstract}
  Channel estimation is a fundamental task in communication systems and is critical for effective demodulation. While most works deal with a simple scenario where the measurements are corrupted by the additive white Gaussian noise (AWGN), this work addresses the more challenging scenario where both AWGN and structured interference coexist. Such conditions arise, for example, when a sonar/radar transmitter and a communication receiver operate simultaneously within the same bandwidth. To ensure accurate channel estimation in these scenarios, the sparsity of 
  the channel in the delay domain and the complicate structure of the interference are jointly 
  exploited. 
  Firstly, the score of the structured interference is learned via a neural network based on the diffusion model (DM), while the channel prior is modeled as a Gaussian distribution, with its variance controlling channel sparsity, 
  similar to the setup of the sparse Bayesian learning (SBL). Then, two efficient posterior sampling 
  methods are proposed to jointly estimate the sparse channel and the interference. Nuisance parameters, such as the variance of the prior are estimated via the expectation maximization (EM) algorithm. The proposed method is termed as DM based SBL (DM-SBL). Numerical simulations demonstrate that DM-SBL significantly outperforms conventional approaches that deal with the AWGN scenario, particularly under low signal-to-interference ratio (SIR) conditions. Beyond channel estimation, DM-SBL also shows promise for addressing other linear inverse problems involving structured interference.
\end{abstract}


\section{Introduction}
Channel estimation, as one of the fundamental issues in communication, has been extensively studied since the inception of digital communication systems. Because the underwater acoustic (UWA) channels commonly exhibit sparsity, compressed sensing (CS) methods have been adopted to address the UWA channel estimation as they can maximally leverage the sparse structure of the channel. \par
Over the past two decades, numerous CS algorithms have been applied to UWA channel estimation, 
including greedy algorithms, Bayesian and optimization based algorithms. 
In \cite{7964794} orthogonal matching pursuit (OMP) is used to estimate the sparse channel 
in UWA OFDM system under impulsive noise. 
For the double-dispersive channel estimation problem, OMP \cite{qu2008estimation}, compressed sampling MP (CoSaMP), sparsity adap
tive MP (SaMP), adaptive step size SaMP (AS-SaMP) \cite{zhang2019efficient} have been proposed. 
In MIMO system, improved SBL (I-SBL) \cite{qin2020bayesian}, forward-reverse SOMP (FRSOMP) \cite{zhou2020exploiting} are proposed to further utilize 
the channel correlation characteristics between different transmitter-receiver pairs. 
It is worth noting that these methods all assume ocean ambient noise to be additive white Gaussian noise (AWGN) or impulsive noise, which is valid in most cases. However, this assumption does not hold in some certain extremal situations. In specific scenarios, such as simultaneous communication during the transmission of detection/imaging signals, or communication near an operating vessel, the interference signals often exhibit distinct structure rather than resembling AWGN. Achieving reliable channel estimation in such scenarios has been a challenging task. Although exhibiting highly structured characteristics, it is almost infeasible to obtain the analytical probability density function (PDF) or the underlying structure of the interference signals. \par
Recently, diffusion models (DMs) has gained remarkable success in the field of unconditional data generation, including audio, image and video data. Although DMs were initially developed for unconditional data generation, they are increasingly being applied in the field of conditional data generation. In short, let $\{\mathbf{x}_1, \mathbf{x}_2, \dots, \mathbf{x}_N\}$ denote 
a dataset of size $N$, whose data points are independent and identically distributed (i.i.d.) samples from 
an underlying data distribution $p_{\rm data}(\mathbf{x})$ \cite{song2022learning}, i.e., 
\begin{equation}
  \{\mathbf{x}_1, \mathbf{x}_2, \dots, \mathbf{x}_N\} \mathop{\sim}\limits^{\rm i.i.d} p_{\rm data}(\mathbf{x}).
\end{equation}
Unconditional data generation means generating a new data point $\mathbf{x}_{N+1}$ from the distribution $p_{\rm data}(\mathbf{x})$. 
Now let $\mathcal{A}$ denote some linear or nonlinear operation and suppose there is an observation $\mathbf{y}$ which 
is obtained by $\mathbf{y} = \mathcal{A}(\mathbf{x}_{N+1})$. The unconditional sampling means to inference 
$\mathbf{x}_{N+1}$ with $\mathbf{y}$ while ensuring that $\mathbf{x}_{N+1} \sim p_{\rm data}(\mathbf{x})$. \par

As a very popular method, conditional sampling via DM has been used in channel estimation, since estimating 
channel from the received signal can be viewed as solving a linear inverse problem. 
In \cite{arvinte2022mimo}, annealed Langevin dynamics is used to estimate the wireless multiple-input multiple-output (MIMO) 
channel, in \cite{zilberstein2024solving} annealed higher-order Langevin dynamics is used to perform 
MIMO channel estimation and solve other linear inverse problem such as Gaussian deblurring, inpainting and super-resolution. 
These methods require training a score-based model for the channel, and they assume that the noise in the models are AWGN. 
As for solving linear inverse problems under structured interference, \cite{stevens2023removing} learns score models for both 
the noise and desired signal, both of them are estimated simultaneously by conditional sampling. Compared to other methods that assume AWGN, this approach addresses the channel estimation under structured interference, meanwhile exploiting the sparse structure of the channel. 
Other approaches, such as \cite{jayashankar2024score}, use score-based diffusion to separate the interesting 
signal from multiple independent sources, then perform data demodulation. 
It is worth noting that using neural networks to 
obtain the scores of the underlying signal enhances the estimation accuracy, while in some applications, using an analytical prior to model the signal of interest allows the algorithm to quickly adapt to various scenarios. 
\par

Motivated by the advantages of DMs for conditional sampling on signals without an analytical prior distribution, a score-based model for the structured interference is learned. Given the sparsity of the channel in delay domain and inspired by SBL, the channels are modeled with a Gaussian distribution with its variance controlling the sparsity of the channel. 
The interference and channel are estimated simultaneously through the conditional sampling process, and the proposed method is named as DM based SBL (DM-SBL). Nuisance parameters involving the prior of channel are updated at each time step via the expectation maximization (EM) algorithm. In summary, our Contributions are as follows:
\begin{itemize}
  \item [1)] A sparse channel estimation method based on score-based DM demonstrating superior performance under structured interference is proposed, compared to classical methods that cope with the AWGN environment. 
  \item[2)] Considering that the prior information of the channel can vary significantly across different environments, the proposed method does not require learning the score of the channel. The nuisance parameters that characterize the prior of the channel are learned via the EM algorithm. This distinguishes the DM-SBL from other diffusion model-based methods, and ensure that DM-SBL is likely to work in the sparse channel setting. 
  \item[3)] The derivations and implement details are provided, and two approximations, i.e., diffusion model based posterior sampling (DMPS) \cite{meng2022diffusion} and pseudoinverse-guided diffusion models ($\Pi$GDM) \cite{song2023pseudoinverseguided}, are employed to approximate the noise-perturbed likelihood. Moreover, except for the score computation by the neural network, where the real and imaginary parts are processed separately, all other operations are performed directly in the complex domain, which reduces the computation complexity when compared with operations evaluating in the real domain.
\end{itemize}

The remainder of this paper is organized as follows: Section \ref{Problem} introduces the problem formulation and 
the background of the score-based diffusion model. Section \ref{method} details how the proposed method works. 
Numerical simulations are presented in Section \ref{numerical}. 
Finally, we draw conclusions in Section \ref{conclu}. 

\textbf{Notation.} The boldfaced letters $\mathbf{x}$, $\mathbf{X}$ denote vectors and matrices, respectively. 
$\mathbf{x}(i)$ denotes the $i$-th element of vector $\mathbf{x}$. $(\cdot)^{\rm T}$, $(\cdot)^{\rm H}$ and 
$(\cdot)^*$ represent the transpose, Hermitian and conjugate, respectively. 
We use $\nabla_{\mathbf{z}^*}f(\mathbf{z}, \mathbf{z}^*)$ to denote the gradient of a function $f$ with 
respect to the vector $\mathbf{z}^*$. $\mathcal{N}(\mathbf{x}; \boldsymbol{\mu}, \mathbf{\Sigma})$ and $\mathcal{CN}(\mathbf{x}; \boldsymbol{\mu}, \mathbf{\Sigma})$ denote the Gaussian distribution and complex Gaussian distribution for random variable $\mathbf{x}$ with mean $\boldsymbol{\mu}$ and covariance $\mathbf{\Sigma}$. For a matrix $\boldsymbol\Sigma$, $|\boldsymbol\Sigma|$ denotes its determinant, while for a vector $\mathbf x$, $|\mathbf x|$ and $\mathbf{x}^{\odot 2}$ denotes its elementwise modulus and square, respectively.

\section{Problem Formulation and Preliminaries}\label{Problem}
In this section, the formulation of channel estimation under structured interference is presented. 
Subsequently, the background of score-based diffusion model is provided.
\subsection{Channel Estimation under Structured Noise}
Consider a general input-output model of a single-carrier communication system, where the 
measurements are corrupted by both 
structured interference $\mathbf{n}$ and AWGN $\boldsymbol{\epsilon}$. 
During the training phase, $N$ pilot symbols $\bar{\mathbf{x}}$ are transmitted, the length of the virtual channel $\mathbf{h}$ is $L$. The received signal is given by 
\begin{equation}\label{sysmodel}
  \mathbf{y}=\mathbf{A}\mathbf{h} + \mathbf{n} + \boldsymbol{\epsilon},
\end{equation}
where $\mathbf{y} \in \mathbb{C}^{M}$ and $ M \triangleq N-L+1 $. $\mathbf{A} \in \mathbb{C}^{M\times L}$ is a Toeplitz matrix whose first column and row are $[\bar{x}_{L-1}, \bar{x}_{L}, \dots, \bar{x}_{N-1}]^{\rm T}$ and 
$[\bar{x}_{L-1}, \bar{x}_{L-2}, \dots, \bar{x}_0]$, respectively. $\mathbf{n} \in \mathbb{C}^{M}$ is the structured interference, which might be 
actively transmitted detection/imaging signals in the form of continue wave (CW), linear frequency modulation (LFM), hyperbolic frequency modulation (HFM), or highly correlated and relatively structured ship propulsion noise. 
$\boldsymbol{\epsilon}\sim \mathcal{CN}(\boldsymbol{\epsilon}; \mathbf{0}, \sigma^2_y \mathbf{I}_M)$ is the AWGN vector. 
$\mathbf{h}$ here is a sparse vector. Similar to the idea of SBL, we model its prior as a zero-mean Gaussian distribution, with its variance controlled by the variable $\boldsymbol{\gamma}$, i.e., 
\begin{equation}\label{h0}
  p(\mathbf{h}; \boldsymbol{\gamma}) = \mathcal{CN}(\mathbf{h}; \mathbf{0}, {\rm diag}(\boldsymbol{\gamma})).
\end{equation}
\subsection{Score-based diffusion model}
Score-based generative models \cite{song2019generative, song2020improved} and 
diffusion probabilistic generative models \cite{ho2020denoising,song2021denoising}
have achieved significant breakthroughs in the field of image generation. In \cite{song2020score}, a unified framework which encapsulates both approaches by expressing
diffusion as a continuous-time process through stochastic 
differential equations (SDE) is proposed. This framework also facilitates reverse sampling at arbitrary time intervals, which 
greatly improves sampling efficiency compared to traditional methods such as denoising diffusion probabilistic models (DDPM) \cite{ho2020denoising}. 
\par
The goal of the forward process of SDE is to transform data into a simple Gaussian distribution 
with zero mean and unit variance. Denote the noiseless data as $\mathbf{x}_0 \in \mathbb{R}^{d}$, the forward process of SDE can be described as 
\begin{equation}
  {\rm d}\mathbf{x}_t = \mathbf{f}(\mathbf{x}_t, t){\rm d}t + g(t){\rm d}\mathbf{w},
\end{equation}
where $\mathbf{w}$ is the standard Wiener process, $\mathbf{f}(\cdot, t): \mathbb{R}^{d} \rightarrow \mathbb{R}^{d}$ and $g(t): \mathbb{R} \rightarrow \mathbb{R}$ 
are the drift and diffusion coefficients, respectively. Specifically, when adopting 
variance preserving (VP) SDE, we have $\mathbf{f}(\mathbf{x}_t, t) = -\frac{1}{2}\beta(t)\mathbf{x}_t$ and $g(t)=\sqrt{\beta(t)}$, where 
\begin{equation}
  \beta(t) = \beta_{\rm min} + t(\beta_{\rm max} - \beta_{\rm min}).
\end{equation}
Subsequently, we have the perturbation kernels 
\begin{equation}
  p(\mathbf{x}_t|\mathbf{x}_0)=\mathcal{N}(\mathbf{x}_t ; \alpha(t)\mathbf{x}_0, 
  (1 - \alpha^2(t)) \mathbf{I}_{d}), \ t \in [0, 1]
\end{equation}
and 
\begin{equation}
  \alpha(t)=\exp \left(
  -\frac{1}{4} t (\beta(t) - \beta_{\rm min})
  \right).
\end{equation}
 \par
In the reverse process of SDE, we obtain samples $\mathbf{x}_0$ by starting sampling from 
$\mathbf{x}_1$. The reverse-time SDE is given by \cite{song2020score,anderson1982reverse}
\begin{equation}
  {\rm d}\mathbf{x}_t = [\mathbf{f}(\mathbf{x}_t, t) - g^2(t) \nabla_{\mathbf{x}_t} \log p(\mathbf{x}_t) ]{\rm d}t + g(t){\rm d}\bar{\mathbf{w}},
\end{equation}
where $\bar{\mathbf{w}}$ is the standard Wiener process in the reverse direction. 
The gradient of the log-likelihood of $\mathbf{x}_t$ with respect to itself is called 
score function, and it is estimated by training a score-based model $\mathbf{s}_{\boldsymbol{\theta}}(\mathbf{x}_t, t)$ on samples with
score matching, i.e., 
\begin{small}
\begin{equation}
  \boldsymbol{\theta}^{*} = \underset{\boldsymbol{\theta}}{\operatorname{argmin}}
  \ {\rm E}_{t} 
  \left\{
    {\rm E}_{\mathbf{x}_0} {\rm E}_{\mathbf{x}_t|\mathbf{x}_0}
    \left[
      \| \mathbf{s}_{\boldsymbol{\theta}}(\mathbf{x}_t, t) 
      - \nabla_{\mathbf{x}_t} \log p(\mathbf{x}_t|\mathbf{x}_0)
      \Vert_2^2 
    \right]
  \right\}.
\end{equation}
\end{small}
In practice, $t$ is randomly sampled from a uniform distribution over the interval $[0, 1]$, 
$\mathbf{x}_0 \sim p(\mathbf{x}_0)$ is sampled from the training dataset and $\mathbf{x}_t \sim p(\mathbf{x}_t|\mathbf{x}_0)$. After obtaining the score, the reverse process of SDE can be done by numerical samplers such as the state-of-the-art Predictor-Corrector sampler \cite{song2020score}. 
Note that in score-based generative models, the variable $\mathbf{x}$ should be real-valued, whereas in our problem all variables are complex-valued. This issue can be addressed by separately handling the real and imaginary parts in both the  
forward and reverse process. 
\section{DM-SBL}\label{method}
\begin{figure}[htpb]
  \centering
  \includegraphics[width=3in]{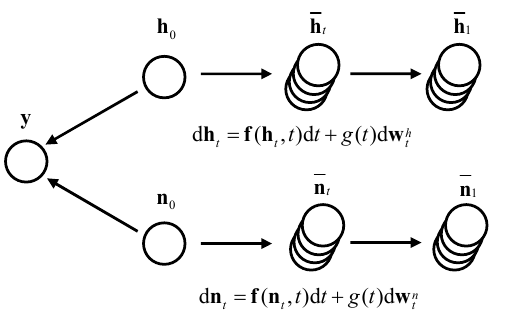}
  \caption{Probabilistic graph of the forward process and system model.}
  \label{forward}
\end{figure}

In our problem, estimating $\mathbf{h}$ under structured noise $\mathbf{n}$ is solved by 
the reverse process of SDE. 
Most current methods that apply diffusion models to solve inverse problems rely on neural networks to 
learn the score of $\mathbf{h}$. However, in our method, the prior of $\mathbf{h}$ contains parameters 
$\boldsymbol{\gamma}$ that needs to be estimated, which means that $\mathbf{h}$ lacks accurate prior 
information at the initial sampling stage. Therefore, we perform the reverse sampling process parallelly on $K$ samples of both $\mathbf{h}$ and $\mathbf{n}$ simultaneously. 
Numerical simulations have shown that this strategy benefits the DM-SBL.  
We denote the $i$-th sample of $\mathbf{h}$ at time instance $t$ as $\mathbf{h}_t^{(i)}$, similarly, $\mathbf{n}_t^{(j)}$ represents the $j$-th sample of $\mathbf{n}$ at time instance $t$. 
For any $\mathbf{h}_t^{(i)}$ and $\mathbf{n}_t^{(j)}$, their conditional probability density function (PDF) are determined by the following perturbation kernels, i.e.,
\begin{equation}\label{ht_h0}
  p(\mathbf{h}_t^{(i)}|\mathbf{h}_0)=\mathcal{CN}(\mathbf{h}_t^{(i)}; \alpha(t) \mathbf{h}_0, 2(1 - \alpha^2(t))\mathbf{I}_L)
\end{equation}
and 
\begin{equation}
  p(\mathbf{n}_t^{(j)}|\mathbf{n}_0)=\mathcal{CN}(\mathbf{n}_t^{(j)}; \alpha(t) \mathbf{n}_0, 2(1 - \alpha^2(t))\mathbf{I}_M).
\end{equation}
The relationship among these variables are shown in Fig. \ref{forward} in the form of probabilistic graph, where $\bar{\mathbf{h}}_t$ and $\bar{\mathbf{n}}_t$ are used to represent the collection of all samples, i.e., $\{\mathbf{h}_t^{(i)}\}_{i=1}^{K}$ and $\{\mathbf{n}_t^{(j)}\}_{j=1}^{K}$, respectively. 
\par
For most existing works which use only one sample of $\mathbf{h}$ and $\mathbf{n}$, the following SDE
\begin{equation}
  \begin{aligned}
    {\rm d}(\mathbf{h}_t, \mathbf{n}_t) &= \left[
    \mathbf{f}(\mathbf{h}_t, \mathbf{n}_t, t)-\right.\\
    &\left.g^2(t) \nabla_{\mathbf{h}_t^*, \mathbf{n}_t^*} \log p(\mathbf{h}_t, \mathbf{n}_t|\mathbf{y})
    \right]{\rm d}t + g(t){\rm d}\bar{\mathbf{w}}
  \end{aligned}   
\end{equation}
is analyzed \cite{stevens2023removing,chung2022diffusion}, where $p(\mathbf{h}_t, \mathbf{n}_t|\mathbf{y})$ is the joint posterior distribution of the 
perturbed channel and interference. 
Now considering the case of using $K$ samples of $\mathbf{h}$ and $\mathbf{n}$, as shown in 
Fig. \ref{forward}, $p(\mathbf{h}_t^{(i)}, \mathbf{n}_t^{(j)}|\mathbf{y})$ is intractable in general. 
Here, we use an implementation trick \cite{jayashankar2024score,dhariwal2021diffusion} to factor $p(\bar{\mathbf{h}}_t, \bar{\mathbf{n}}_t|\mathbf{y}) $ as 
\begin{equation}
  \begin{aligned}
    p(\bar{\mathbf{h}}_t, \bar{\mathbf{n}}_t|\mathbf{y}) 
    & = \prod_{(i,j)\in \mathcal{K}\times\mathcal{K}}
    p(\mathbf{h}_t^{(i)}, \mathbf{n}_t^{(j)}|\mathbf{y})^{\frac{1}{K}}\\
    &= \prod_{(i,j)\in \mathcal{K}\times\mathcal{K}}
    \left[
      p(\mathbf{y}|\mathbf{h}_t^{(i)}, \mathbf{n}_t^{(j)})^{\frac{1}{K}} 
      p(\mathbf{h}_t^{(i)}, \mathbf{n}_t^{(j)})^{\frac{1}{K}}
    \right]\\
    & = \prod_{(i,j)\in \mathcal{K}\times\mathcal{K}}
    \left[
      p(\mathbf{y} | \mathbf{h}_t^{(i)}, \mathbf{n}_t^{(j)})^{\frac{1}{K}}
      p(\mathbf{h}_t^{(i)})^{\frac{1}{K}}
      p(\mathbf{n}_t^{(j)})^{\frac{1}{K}}
    \right],
  \end{aligned}
\end{equation}
where $\mathcal{K} \triangleq \{1, 2, \dots, K\}$. Therefore, the reverse process of SDE corresponding 
to the proposed setup is 
\begin{equation}
  \begin{aligned}
    {\rm d}(\mathbf{h}_t^{(i)}, \mathbf{n}_t^{(j)}) &= \left[
      \mathbf{f}(\mathbf{h}_t^{(i)}, \mathbf{n}_t^{(j)}, t)-
      g^2(t) \nabla_{\mathbf{h}_t^{(i)*}, \mathbf{n}_t^{(j)*}} \log p(\bar{\mathbf{h}}_t, \bar{\mathbf{n}}_t|\mathbf{y})
      \right]{\rm d}t + g(t){\rm d}\bar{\mathbf{w}}
  \end{aligned}  
\end{equation}
According to \cite{stevens2023removing}, we can construct two separate diffusion processes. 
The conditional posterior score is separated into two parts, i.e.,
\begin{equation}\label{posterior_grad}
  \left\{
\begin{aligned}
\nabla_{\mathbf{h}_t^{(i)*}} \log p(\bar{\mathbf{h}}_t, \bar{\mathbf{n}}_t | \mathbf{y}) &\simeq \nabla_{\mathbf{h}_t^{(i)*}} \log p(\mathbf{h}_t^{(i)}) + \frac{1}{K}\sum_{j=1}^{K} \nabla_{\mathbf{h}_t^{(i)*}} \log p(\mathbf{y}|\mathbf{h}_t^{(i)}, \mathbf{n}_t^{(j)}) \\
\nabla_{\mathbf{n}_t^{(j)*}} \log p(\bar{\mathbf{h}}_t, \bar{\mathbf{n}}_t | \mathbf{y}) &\simeq  \nabla_{\mathbf{n}_t^{(j)*}} \log p(\mathbf{n}_t^{(j)}) + \frac{1}{K} \sum_{i=1}^{K} \nabla_{\mathbf{n}_t^{(j)*}} \log p(\mathbf{y}|\mathbf{h}_t^{(i)}, \mathbf{n}_t^{(j)}) \\
\end{aligned}.
  \right.
\end{equation}
It can be seen that for any $\mathbf{h}_t^{(i)}$, they are entangled with all samples in $\bar{\mathbf{n}}_t$ 
through the likelihood $p(\mathbf{y}|\mathbf{h}_t^{(i)}, \mathbf{n}_t^{(j)})$, and vice versa. 
Now we analyze the four terms in (\ref{posterior_grad}). Since the structured noise $\mathbf{n}$ is arbitrarily sampled from a large set of signals, it is impossible to obtain an analytical expression for its PDF, therefore, its score $\nabla_{\mathbf{n}_t^{(j)*}} \log p(\mathbf{n}_t^{(j)})$ can only be 
obtained by training a score-based model denoted by $\mathbf{s}_{\boldsymbol{\theta}}(\mathbf{n}_t^{(j)}, t)$. As for $\nabla_{\mathbf{h}_t^{(i)*}} \log p(\mathbf{h}_t^{(i)})$, since we have both $p(\mathbf{h}_t^{(i)}| \mathbf{h}_0)$ in (\ref{ht_h0}) and $p(\mathbf{h}_0; \boldsymbol{\gamma})$ in (\ref{h0}), this score can be obtained analytically. $p(\mathbf{h}_t^{(i)}; \boldsymbol{\gamma})$ can be obtained as 
\begin{equation}
  \begin{aligned}
    p(\mathbf{h}_t^{(i)}; \boldsymbol{\gamma}) &= \int \mathcal{CN}\left(
    \mathbf{h}_0; \frac{1}{\alpha(t)}\mathbf{h}_t^{(i)}, \frac{2(1 - \alpha^2(t))}{\alpha^2(t)}
    \mathbf{I}_{L}
    \right)\times \mathcal{CN}(\mathbf{h}_0; \mathbf{0}, {\rm diag}(\boldsymbol{\gamma})) {\rm d} \mathbf{h}_0 \\
    &\propto \mathcal{CN}\left(
      \mathbf{h}_t^{(i)}; \mathbf{0}, 2(1 - \alpha^2(t))\mathbf{I}_L + \alpha^2(t) {\rm diag}(\boldsymbol{\gamma})
    \right).
  \end{aligned}
\end{equation}
For brevity, we denote 
\begin{equation}\label{sigmaht}
  \mathbf{\Sigma}_{h, t} = 2(1 - \alpha^2(t))\mathbf{I}_L + \alpha^2(t) {\rm diag}(\boldsymbol{\gamma}),
\end{equation}
and $\log p(\mathbf{h}_t^{(i)})$ can be expressed as
\begin{equation}
  \log p(\mathbf{h}_t^{(i)}; \boldsymbol{\gamma}) = -(\mathbf{h}_t^{(i)})^{\rm H}
  \mathbf{\Sigma}_{h, t}^{-1} \mathbf{h}_t^{(i)} - \log |\mathbf{\Sigma}_{h, t}| + {\rm const},
\end{equation}
where ${\rm const}$ represents a term that is independent of $\mathbf{h}_t$ and $\boldsymbol{\gamma}$. Consequently, $ \nabla_{\mathbf{h}_t^{(i)*}} \log p(\mathbf{h}_t^{(i)})$ is 
calculated as 
\begin{equation}
  \nabla_{\mathbf{h}_t^*} \log p(\mathbf{h}_t^{(i)}) = -\mathbf{\Sigma}_{h, t}^{-1} \mathbf{h}_t^{(i)}.
\end{equation}
As for the noise-perturbed likelihood $p(\mathbf{y}|\mathbf{h}_t^{(i)}, \mathbf{n}_t^{(j)})$, multiple 
independent works have proposed different approximations for it, including DMPS \cite{meng2022diffusion}, $\Pi$GDM \cite{song2023pseudoinverseguided} and Diffusion Posterior Sampling (DPS) \cite{chung2022diffusion}. 
In this work, DMPS and $\Pi$GDM are chosen as our approximation methods, which are detailed in 
the following two subsections. 
Note that in 
the following two subsections,  we temporally drop the notation $i$ and $j$ on $\mathbf{h}_t$ and $\mathbf{n}_t$ for 
brevity. 

\begin{table*}[h!]
  \caption{Approximated noise-perturbed likelihood scores}
  \centering
  \renewcommand\arraystretch{1.5}
  \label{like}
  \begin{tabular}{|m{1cm}<{\centering}|m{5cm}<{\centering}|m{5cm}<{\centering}|}
  \hline
      ~ & $\nabla_{\mathbf{h}_t^*} \log p(\mathbf{y}|\mathbf{h}_t, \mathbf{n}_t)$ & $\nabla_{\mathbf{n}_t^*} \log p(\mathbf{y}|\mathbf{h}_t, \mathbf{n}_t)$ \\ \hline
      DMPS & $\frac{1}{\alpha(t)} \mathbf{A}^{\rm H} \mathbf{\Sigma}_{y,t}^{-1} 
      \left(
        \mathbf{y} - \frac{1}{\alpha(t)} \mathbf{A}\mathbf{h}_t - \frac{1}{\alpha(t)} \mathbf{n}_t
      \right)$ & $\frac{1}{\alpha(t)} \mathbf{\Sigma}_{y,t}^{-1} 
      \left(
        \mathbf{y} - \frac{1}{\alpha(t)} \mathbf{A}\mathbf{h}_t - \frac{1}{\alpha(t)} \mathbf{n}_t
      \right)$ \\ \hline
      $\Pi$GDM & $(\nabla_{\mathbf{h}_t} \widehat{\mathbf{h}}_t)^* \mathbf{A}^{\rm H} \mathbf{\Sigma}_{y, t}^{-1}
  (\mathbf{y} - \mathbf{A}\widehat{\mathbf{h}}_t - \widehat{\mathbf{n}}_t)$ & $(\nabla_{\mathbf{n}_t} \widehat{\mathbf{n}}_t)^*  \mathbf{\Sigma}_{y, t}^{-1}
  (\mathbf{y} - \mathbf{A}\widehat{\mathbf{h}}_t - \widehat{\mathbf{n}}_t)$ \\ \hline
  \end{tabular}
\end{table*}

\subsection{Perturbation likelihood using DMPS}
According to DMPS, given the perturbation kernels, $\mathbf{h}_0$ and $\mathbf{n}_0$ can be 
represented as \cite{meng2022diffusion}
\begin{align}
  \mathbf{h}_0 &= \frac{\mathbf{h}_t - \sqrt{2(1 - \alpha^2(t))} \mathbf{w}_h}{\alpha(t)} \label{h0_ht}\\
  \mathbf{n}_0 &= \frac{\mathbf{n}_t - \sqrt{2(1 - \alpha^2(t))} \mathbf{w}_n}{\alpha(t)}, \label{n0_nt}
\end{align}
where $\mathbf{w}_h$ and $\mathbf{w}_n$ are AWGN with unit variance. 
Consequently, substituting (\ref{h0_ht}) and (\ref{n0_nt}) into (\ref{sysmodel}), the system model becomes 
\begin{equation}
  \begin{aligned}
    \mathbf{y} &= \frac{1}{\alpha(t)} \mathbf{A} \mathbf{h}_t + 
    \frac{1}{\alpha(t)}\mathbf{n}_t - \frac{\sqrt{2(1 - \alpha^2(t))}}{\alpha(t)}\mathbf{A}
    \mathbf{w}_h-\frac{\sqrt{2(1 - \alpha^2(t))}}{\alpha(t)}\mathbf{w}_n + \boldsymbol{\epsilon}.
  \end{aligned}
\end{equation}
As $\mathbf{w}_h$, $\mathbf{w}_n$ and $\boldsymbol{\epsilon}$ are independent of each other, we can obtain the approximated noise perturbed likelihood function $p(\mathbf{y}|\mathbf{h}_t, \mathbf{n}_t)$ as
\begin{equation}
  \begin{aligned}
    p(\mathbf{y}|\mathbf{h}_t, \mathbf{n}_t) = &\mathcal{CN}\left(
    \mathbf{y}; \frac{1}{\alpha(t)} \mathbf{A}\mathbf{h}_t + \frac{1}{\alpha(t)} \mathbf{n}_t, \frac{2(1-\alpha^2(t))}{\alpha^2(t)}(\mathbf{A}\mathbf{A}^{\rm H} + \mathbf{I}_M) + \sigma^2_y \mathbf{I}_{M} 
  \right)
  \end{aligned}
\end{equation}
For brevity, we denote $\mathbf{\Sigma}_{y,t} \triangleq \frac{2(1-\alpha^2(t))}{\alpha^2(t)}(\mathbf{A}\mathbf{A}^{\rm H} + \mathbf{I}_M) + \sigma^2_y \mathbf{I}_{M}$
Therefore, the noise-perturbed likelihood scores are calculated as
\begin{equation}\label{DMPS}
  \left\{
\begin{aligned}
&\nabla_{\mathbf{h}_t^{*}} \log p(\mathbf{y} | \mathbf{h}_t, \mathbf{n}_t) = \frac{1}{\alpha(t)} \mathbf{A}^{\rm H} \mathbf{\Sigma}_{y,t}^{-1} 
\left(
  \mathbf{y} - \frac{1}{\alpha(t)} \mathbf{A}\mathbf{h}_t 
- \frac{1}{\alpha(t)} \mathbf{n}_t
\right) \\
&\nabla_{\mathbf{n}_t^*} \log p(\mathbf{y} | \mathbf{h}_t, \mathbf{n}_t) = \frac{1}{\alpha(t)} \mathbf{\Sigma}_{y,t}^{-1} 
\left(
  \mathbf{y} - \frac{1}{\alpha(t)} \mathbf{A}\mathbf{h}_t - \frac{1}{\alpha(t)} \mathbf{n}_t
\right) \\
\end{aligned}.
  \right.
\end{equation}
Substituting (\ref{DMPS}) into (\ref{posterior_grad}), we obtain 
\begin{equation}\label{DMPS_post}
  \left\{
\begin{aligned}
&\nabla_{\mathbf{h}_t^{(i)*}} \log p(\bar{\mathbf{h}}_t, \bar{\mathbf{n}}_t | \mathbf{y}) \simeq \nabla_{\mathbf{h}_t^{(i)*}} \log p(\mathbf{h}_t^{(i)}) + \frac{1}{\alpha(t)}\mathbf{A}^{\rm H} \mathbf{\Sigma}_{y,t}^{-1}
\left(\mathbf{y} - \frac{1}{\alpha(t)}\mathbf{A}^{\rm H} \mathbf{h}_t^{(i)} - \frac{1}{\alpha(t) K }\sum_{j=1}^K \mathbf{n}_t^{(j)} \right) \\
&\nabla_{\mathbf{n}_t^{(j)*}} \log p(\bar{\mathbf{h}}_t, \bar{\mathbf{n}}_t | \mathbf{y}) \simeq  \nabla_{\mathbf{n}_t^{(j)*}} \log p(\mathbf{n}_t^{(j)}) + \frac{1}{\alpha(t)} \mathbf{\Sigma}_{y,t}^{-1} 
\left(
  \mathbf{y} - \frac{1}{\alpha(t) K} \mathbf{A}^{\rm H} \sum_{i=1}^{K}\mathbf{h}_t^{(i)}
  -\frac{1}{\alpha(t)}\mathbf{n}_t^{(j)}
\right)\\
\end{aligned}.
  \right.
\end{equation}

\subsection{Perturbation likelihood using $\Pi$GDM}
According to $\Pi$GDM, we approximate $p(\mathbf{n}_0|\mathbf{n}_t)$ by a Gaussian distribution 
$\mathcal{CN}(\mathbf{n}_0; \widehat{\mathbf{n}}_t, r_t^2 \mathbf{I}_M)$, where $\widehat{\mathbf{n}}_t$ is obtained by Tweedie estimator \cite{kotz2012breakthroughs} \footnote{${\rm E}_{\mu|\theta}[\mu]=\theta + \frac{{\rm d} \ln p(\theta|\mu)}{{\rm d} \theta^*}$ for $p(\theta|\mu) = \mathcal{CN}(\theta; \mu, \sigma^2)$} given by 
\begin{equation}
  \widehat{\mathbf{n}}_t = \frac{1}{\alpha(t)}{\rm E}[\alpha(t)\mathbf{n}_0|\mathbf{n}_t]=
  \frac{\mathbf{n}_t + 2(1 - \alpha^2(t)) \nabla_{\mathbf{n}_t^{*}} \log p(\mathbf{n}_t)}{\alpha(t)}.
\end{equation} 
Here $r_t = \sqrt{2(1 - \alpha^2(t))}$ is an empirical parameter, 
$p(\mathbf{h}_0|\mathbf{h}_t)$ can be also obtained as $\mathcal{CN}(\mathbf{h}_0; \widehat{\mathbf{h}}_t, r^2_t \mathbf{I}_L)$ by the same step. 

Similar to that of DMPS, the noise perturbed likelihood can be calculated as 
\begin{equation}
  p(\mathbf{y}|\mathbf{h}_t, \mathbf{n}_t)=\mathcal{CN}\left(
    \mathbf{y}; \mathbf{A}\widehat{\mathbf{h}}_t + \widehat{\mathbf{n}}_t, 
    r_t^2(\mathbf{A} \mathbf{A}^{\rm H} + \mathbf{I}_M) + \sigma_y^2 \mathbf{I}_M
  \right).
\end{equation}
Again, for brevity, we denote $\mathbf{\Sigma}_{y,t} \triangleq r_t^2(\mathbf{A} \mathbf{A}^{\rm H} + \mathbf{I}_M) + \sigma_y^2 \mathbf{I}_M$, therefore, we have 
\begin{equation}\label{PGDM}
  \left\{
\begin{aligned}
  \nabla_{\mathbf{h}_t^*} \log p(\mathbf{y}|\mathbf{h}_t, \mathbf{n}_t) &= 
  (\nabla_{\mathbf{h}_t} \widehat{\mathbf{h}}_t)^* \mathbf{A}^{\rm H} \mathbf{\Sigma}_{y, t}^{-1}
  (\mathbf{y} - \mathbf{A}\widehat{\mathbf{h}}_t - \widehat{\mathbf{n}}_t) \\
  \nabla_{\mathbf{n}_t^*} \log p(\mathbf{y}|\mathbf{h}_t, \mathbf{n}_t) &= 
  (\nabla_{\mathbf{n}_t} \widehat{\mathbf{n}}_t)^*  \mathbf{\Sigma}_{y, t}^{-1}
  (\mathbf{y} - \mathbf{A}\widehat{\mathbf{h}}_t - \widehat{\mathbf{n}}_t)
\end{aligned}
\right.,
\end{equation}
where $\nabla_{\mathbf{n}_t} \widehat{\mathbf{n}}_t$ can be obtained by automatic differentiation methods of pytorch or other 
prevailing machine learning framework, and $\nabla_{\mathbf{h}_t} \widehat{\mathbf{h}}_t$ can be computed analytically as
\begin{equation}
  \nabla_{\mathbf{h}_t} \widehat{\mathbf{h}}_t=\frac{\mathbf{I}_M + 2(1 - \alpha^2(t))\boldsymbol{\Sigma}_{h,t}^{-1}}{\alpha(t)},
\end{equation}
where $\boldsymbol{\Sigma}_{h,t}$ is defined by (\ref{sigmaht}). 
Again, substituting (\ref{PGDM}) into (\ref{posterior_grad}), we have
\begin{equation}
\left\{
\begin{aligned}
&\nabla_{\mathbf{h}_t^{(i)*}} \log p(\bar{\mathbf{h}}_t, \bar{\mathbf{n}}_t | \mathbf{y}) \simeq \nabla_{\mathbf{h}_t^{(i)*}} \log p(\mathbf{h}_t^{(i)}) + (\nabla_{\mathbf{h}_t^{(i)}} \widehat{\mathbf{h}}_t^{(i)})^* \mathbf{A}^{\rm H} \mathbf{\Sigma}_{y, t}^{-1}
  \left(\mathbf{y} - \mathbf{A}\widehat{\mathbf{h}}_t^{(i)} - \frac{1}{K}\sum_{j=1}^{K} \widehat{\mathbf{n}}_t^{(j)}\right) \\
&\nabla_{\mathbf{n}_t^{(j)*}} \log p(\bar{\mathbf{h}}_t, \bar{\mathbf{n}}_t | \mathbf{y}) \simeq  \nabla_{\mathbf{n}_t^{(j)*}} \log p(\mathbf{n}_t^{(j)}) +(\nabla_{\mathbf{n}_t^{(j)}} \widehat{\mathbf{n}}_t^{(j)})^*  \mathbf{\Sigma}_{y, t}^{-1}
  \left(\mathbf{y} - \frac{1}{K} \sum_{i=1}^{K} \mathbf{A}\widehat{\mathbf{h}}_t^{(i)} - \widehat{\mathbf{n}}_t^{(j)}\right) \\
\end{aligned}.
  \right.
\end{equation}

We list the approximated noise-perturbed likelihood scores in Table \ref{like}, and we use both approximations from DMPS and $\Pi$GDM to design DM-SBL, respectively. 
We term them as DM-SBL (DMPS) and DM-SBL ($\Pi$GDM).

\subsection{Updating $\boldsymbol{\gamma}$}
In SBL, the nusiance parameter $\boldsymbol{\gamma}$ which controls the sparsity of the desired signal $\mathbf h$ is updated via the EM in each iteration. In DM-SBL, we assume that all the samples of $\mathbf{h}$ share a same prior distribution
\begin{equation}
  \mathbf{h}^{(i)} \sim \mathcal{CN}(\mathbf{h}^{(i)}; \mathbf{0}, {\rm diag}(\boldsymbol{\gamma})), \ {\rm for} \
  i=1, 2, \dots , K
\end{equation}
and at each time step $t$, we update $\boldsymbol{\gamma}$ by EM algorithm, i.e.,
\begin{equation}
  \boldsymbol{\gamma}_{\rm new} = \underset{\boldsymbol{\gamma}}{\operatorname{argmax}} \ 
  {\rm E}_{\mathbf{h}_t, \mathbf{n}_t | \mathbf{y}} [\log p(\mathbf{h}_t; \boldsymbol{\gamma})],
\end{equation}
where
\begin{equation}
  \begin{aligned}
    L(\boldsymbol{\gamma}) &\triangleq {\rm E}_{\mathbf{h}_t, \mathbf{n}_t | \mathbf{y}} [\log p(\mathbf{h}_t; \boldsymbol{\gamma})]
    = - {\rm E}[\mathbf{h}_t^{\rm H} \mathbf{\Sigma}_{h,t}^{-1}\mathbf{h}_t] - \log |\mathbf{\Sigma}_{h, t}| \\
    &= - \sum_{l=0}^{L - 1} \frac{{\rm E}[|\mathbf{h}_t(l)|^2]}{2 (1 - \alpha^2(t)) + \alpha^2(t)\boldsymbol{\gamma}(l)}
    - \log \prod_{l=0}^{L - 1} \left(
      2 (1 - \alpha^2(t)) + \alpha^2(t)\boldsymbol{\gamma}(l)
    \right) \\
    &=-\sum_{l=0}^{L - 1} \frac{\boldsymbol{\nu}^h_t(l) + |\widehat{\mathbf{h}}_t(l)|^2}{2 (1 - \alpha^2(t)) + \alpha^2(t)\boldsymbol{\gamma}(l)}- \sum_{l=0}^{L - 1} \log \left(
      2 (1 - \alpha^2(t)) + \alpha^2(t)\boldsymbol{\gamma}(l)
      \right),
  \end{aligned}  
\end{equation}
where $\widehat{\mathbf{h}}_t$ and $\boldsymbol{\nu}^h_t$ are the mean and variance of $\mathbf{h}_t$, respectively. 
Note that in practice, $\mathbf{h}_t$ is obtained by sampling. Therefore, $p(\mathbf{h}_t, \mathbf{n}_t|\mathbf{y})$ does not 
have an analytical expression. To obtain $\widehat{\mathbf{h}}_t$ and $\boldsymbol{\nu}^h_t$, in the sampling process, multiple 
samples of $\mathbf{h}$ are computed simultaneously, therefore, $\widehat{\mathbf{h}}_t$ and $\boldsymbol{\nu}^h_t$ are 
approximated using their corresponding sample moments, i.e.,
\begin{equation}
    \begin{aligned}
        \widehat{\mathbf{h}}_t &= \frac{1}{K} \sum_{i=1}^K \mathbf{h}_t^{(i)}\\
        \boldsymbol{\nu}_t^h &= \frac{1}{K} \sum_{i=1}^K |\mathbf{h}_t^{(i)} - \widehat{\mathbf{h}}_t|^{\odot2},
    \end{aligned}
\end{equation}
where $|\mathbf{z}|^{\odot2}$ denotes the elemenwise squared modulus of $\mathbf{z}$.
Although the overall computational load has increased due to this operation, all calculations for $\mathbf{h}_t$ can be executed in parallel at a fixed $t$. Additionally, with the high optimization of modern GPUs for parallel computing, the actual increase in computation time is almost negligible.

\begin{figure*}[h!]
  \centering
  \includegraphics[width=6in]{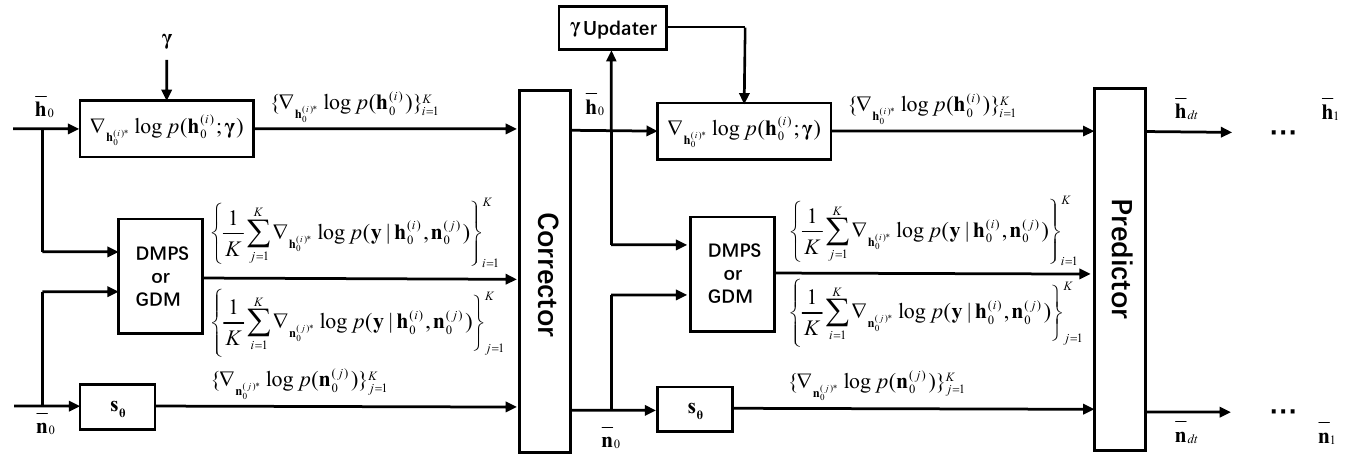} 
  \caption{Iterative joint conditional sampling procedure for DM-SBL.}
  \label{ag1}
\end{figure*}

To maximize $L(\boldsymbol{\gamma})$, we first calculate $\frac{\partial L(\boldsymbol{\gamma}) }{\partial \boldsymbol{\gamma}(l)}$, 
\begin{equation}
  \begin{aligned}
    \frac{\partial L(\boldsymbol{\gamma}) }{\partial \boldsymbol{\gamma}(l)} &= 
    \frac{\alpha^2(t) (\boldsymbol{\nu}^h_t(l) + |\widehat{\mathbf{h}}_t(l)|^2) }{(2(1 - \alpha^2(t)) + \alpha^2(t) \boldsymbol{\gamma}(l))^2}-\frac{\alpha^2(t)}{2(1 - \alpha^2(t)) + \alpha^2(t) \boldsymbol{\gamma}(l)}.
  \end{aligned}
\end{equation}
By letting $\frac{\partial L(\boldsymbol{\gamma}) }{\partial \boldsymbol{\gamma}(l)}$ equal to zero, we can obtain the optimal 
$\boldsymbol{\gamma}_{\rm new}(l)$ as
\begin{equation}
  \boldsymbol{\gamma}_{\rm new}(l) = \frac{1}{\alpha^2(t)}\left(
    \boldsymbol{\nu}_t^h(l) + |\widehat{\mathbf{h}}_t(l)|^2
  \right) - \frac{2(1 - \alpha^2(t))}{\alpha^2(t)}.
\end{equation}
Thereby, the DM-SBL can be summarized in Algorithm \ref{alg1}, where the predictor and corrector sampling algorithm \cite{song2020score} is implemented, the corresponding flow diagram is shown in Fig. \ref{ag1}. 
The structure of the neural network we used for learning the score of the interference is detailed in Appendix \ref{NNAppendix}. To improve the stability of the proposed method, two hyper-parameters $\mu$ and $\kappa$ are introduced to weigh 
the importance of the prior and perturbed likelihood function, i.e., we rewrite (\ref{posterior_grad}) as 
\begin{equation}\label{posterior_grad_prac}
  \left\{
\begin{aligned}
&\nabla_{\mathbf{h}_t^{(i)*}} \log p(\bar{\mathbf{h}}_t, \bar{\mathbf{n}}_t | \mathbf{y}) \simeq \mu \nabla_{\mathbf{h}_t^{(i)*}} \log p(\mathbf{h}_t^{(i)})  + \frac{1}{K}\sum_{j=1}^{K} \nabla_{\mathbf{h}_t^{(i)*}} \log p(\mathbf{y}|\mathbf{h}_t^{(i)}, \mathbf{n}_t^{(j)}) \\
&\nabla_{\mathbf{n}_t^{(j)*}} \log p(\bar{\mathbf{h}}_t, \bar{\mathbf{n}}_t | \mathbf{y}) \simeq  \kappa \nabla_{\mathbf{n}_t^{(j)*}} \log p(\mathbf{n}_t^{(j)}) + \frac{1}{K} \sum_{i=1}^{K} \nabla_{\mathbf{n}_t^{(j)*}} \log p(\mathbf{y}|\mathbf{h}_t^{(i)}, \mathbf{n}_t^{(j)}) \\
\end{aligned}.
  \right.
\end{equation}

\begin{algorithm*}[t!]
  \caption{DM-SBL}
  \renewcommand{\algorithmicrequire}{\textbf{Input:}}
  \renewcommand{\algorithmicensure}{\textbf{Output:}}
  \newcommand{\algorithmicinit}{\textbf{Initialize: }}
  \newcommand{\algorithmicrun}{\textbf{Run: }}
  \begin{algorithmic}[1]\label{alg1}
      \REQUIRE $\mathbf{s}_{\boldsymbol{\theta}}$, $\lambda$, $\kappa$, $\sigma_{y}^2$, $\mu$, $T$
      \ENSURE $\frac{1}{K}\sum_{i=1}^{K}\mathbf{h}_0^{(i)}$ 
      \STATE \algorithmicinit $\mathbf{h}_1^{(i)} \sim \mathcal{CN}(\mathbf{0}, (2(1 - \alpha^2(1))\mathbf{I}_{L})$, 
      $\mathbf{n}_1^{(j)} \sim \mathcal{CN}(\mathbf{0}, (2(1 - \alpha^2(1))\mathbf{I}_{M})$, $\boldsymbol{\gamma} \leftarrow \rho \mathbf{1}_{L}$, ${\rm d} t \leftarrow -\frac{1}{T} $
      \FOR {$i = T - 1$ to 0}
        \STATE $t \leftarrow \frac{i+1}{T}$
        \STATE \textbf{Corrector}
        \STATE $\nabla_{\mathbf{h}_t^{(i)*}} \log p(\bar{\mathbf{h}}_t, \bar{\mathbf{n}}_t | \mathbf{y}) \leftarrow 
        \mu \nabla_{\mathbf{h}_t^{(i)*}} \log p(\mathbf{h}_t^{(i)}) + \frac{1}{K}\sum_{j=1}^{K} \nabla_{\mathbf{h}_t^{(i)*}} \log p(\mathbf{y}|\mathbf{h}_t^{(i)}, \mathbf{n}_t^{(j)})$
        \STATE $\xi_t \leftarrow \nu / \| \nabla_{\mathbf{h}^{(i)*}_t} \log p(\bar{\mathbf{h}}_t, \bar{\mathbf{n}}_t | \mathbf{y}) \Vert_2^2 $        
        \STATE $\mathbf{h}_{t}^{(i)} \leftarrow \mathbf{h}_{t}^{(i)} + \xi_t \nabla_{\mathbf{h}^{(i)*}_t} \log p(\bar{\mathbf{h}}_t, \bar{\mathbf{n}}_t | \mathbf{y})$
        \STATE $\mathbf{z}^{(i)} \sim \mathcal{CN}(\mathbf{0}, 2\mathbf{I}_L)$
        \STATE $\mathbf{h}_{t}^{(i)} \leftarrow \mathbf{h}_{t}^{(i)} + \sqrt{2 \xi_t}\mathbf{z}^{(i)}$
        \STATE $\nabla_{\mathbf{n}_t^{(j)*}} \log p(\bar{\mathbf{h}}_t, \bar{\mathbf{n}}_t | \mathbf{y}) \simeq  \kappa \nabla_{\mathbf{n}_t^{(j)*}} \log p(\mathbf{n}_t^{(j)}) + \frac{1}{K} \sum_{i=1}^{K} \nabla_{\mathbf{n}_t^{(j)*}} \log p(\mathbf{y}|\mathbf{h}_t^{(i)}, \mathbf{n}_t^{(j)})$
        \STATE $\xi_t \leftarrow \nu / \| \nabla_{\mathbf{n}^{(j)*}_t} \log p(\bar{\mathbf{h}}_t, \bar{\mathbf{n}}_t | \mathbf{y}) \Vert_2^2 $        
        \STATE $\mathbf{n}_{t}^{(j)} \leftarrow \mathbf{n}_{t}^{(j)} + \xi_t \nabla_{\mathbf{n}^{(j)*}} \log p(\bar{\mathbf{h}}_t, \bar{\mathbf{n}}_t | \mathbf{y})$
        \STATE $\mathbf{z}^{(j)} \sim \mathcal{CN}(\mathbf{0}, 2\mathbf{I}_M)$
        \STATE $\mathbf{n}_{t}^{(j)} \leftarrow \mathbf{n}_{t}^{(j)} + \sqrt{2 \xi_t}\mathbf{z}^{(j)}$
        \STATE \textbf{Update} $\boldsymbol{\gamma}$
        \STATE $\boldsymbol{\gamma} = \frac{1}{\alpha^2(t)}\left(\boldsymbol{\nu}^h_t + |\widehat{\mathbf{h}}_t|^{\odot 2}\right) - \frac{2(1 - \alpha^2(t))}{\alpha^2(t)}$
        \STATE \textbf{Predictor}
        \STATE $\nabla_{\mathbf{h}_t^{(i)*}} \log p(\bar{\mathbf{h}}_t, \bar{\mathbf{n}}_t | \mathbf{y}) \leftarrow 
        \mu \nabla_{\mathbf{h}_t^{(i)*}} \log p(\mathbf{h}_t^{(i)}) + \frac{1}{K}\sum_{j=1}^{K} \nabla_{\mathbf{h}_t^{(i)*}} \log p(\mathbf{y}|\mathbf{h}_t^{(i)}, \mathbf{n}_t^{(j)})$
        \STATE $\mathbf{h}_{t-{\rm d}t}^{(i)} \leftarrow \mathbf{h}_{t - {\rm d}t}^{(i)} - (\frac{1}{2} \beta(t) \mathbf{h}_{t}^{(i)} 
        + \beta(t) \nabla_{\mathbf{h}^{(i)*}} \log p(\bar{\mathbf{h}}_t, \bar{\mathbf{n}}_t | \mathbf{y})){\rm d}t$
        \STATE $\mathbf{z}^{(i)} \sim \mathcal{CN}(\mathbf{0}, 2\mathbf{I}_{L})$
        \STATE $\mathbf{h}_{t - {\rm d}t}^{(i)} \leftarrow \mathbf{h}_{t - {\rm d}t}^{(i)} + \sqrt{-\beta(t) {\rm d}t} \mathbf{z}^{(i)}$
        \STATE $\nabla_{\mathbf{n}_t^{(j)*}} \log p(\bar{\mathbf{h}}_t, \bar{\mathbf{n}}_t | \mathbf{y}) \simeq  \kappa \nabla_{\mathbf{n}_t^{(j)*}} \log p(\mathbf{n}_t^{(j)}) + \frac{1}{K} \sum_{i=1}^{K} \nabla_{\mathbf{n}_t^{(j)*}} \log p(\mathbf{y}|\mathbf{h}_t^{(i)}, \mathbf{n}_t^{(j)})$
        \STATE $\mathbf{n}_{t-{\rm d}t}^{(j)} \leftarrow \mathbf{n}_{t - {\rm d}t}^{(j)} - (\frac{1}{2} \beta(t) \mathbf{n}_{t}^{(j)}
        + \beta(t) \nabla_{\mathbf{n}^{(j)*}} \log p(\bar{\mathbf{h}}_t, \bar{\mathbf{n}}_t | \mathbf{y})){\rm d}t$
        \STATE $\mathbf{z}^{(j)} \sim \mathcal{CN}(\mathbf{0}, 2\mathbf{I}_{M})$
        \STATE $\mathbf{n}_{t - {\rm d}t}^{(j)} \leftarrow \mathbf{n}_{t - {\rm d}t}^{(j)} + \sqrt{-\beta(t) {\rm d}t} \mathbf{z}^{(j)}$
        \STATE \textbf{Update} $\boldsymbol{\gamma}$
        \STATE $\boldsymbol{\gamma} = \frac{1}{\alpha^2(t)}\left(\boldsymbol{\nu}^h_t + |\widehat{\mathbf{h}}_t|^{\odot 2}\right) - \frac{2(1 - \alpha^2(t))}{\alpha^2(t)}$
      \ENDFOR
  \end{algorithmic}
\end{algorithm*}

\section{Numerical simulation}\label{numerical}

\renewcommand{\arraystretch}{1.2}
\begin{table}[htpb]
  \caption{Parameters used in simulation}
  \centering
  \label{para}
  \begin{tabular}{|c|c|}
  \hline
  Measurement length $M$ & 200 \\
  \hline
  Assumed channel length $L$ & 200, 300\\
  \hline
  Number of channel paths $p_0$ & 10, 15 \\
  \hline
  Bandwidth of LFM $B$ & 1 kHz \\
  \hline
  Duration of LFM $T_{\rm LFM}$ & 2 s \\
  \hline
  Symbol rate $f_{\rm sym}$ & 4 kHz \\
  \hline
  Modulation scheme of pilot & BPSK \\
  \hline
  \end{tabular}
\end{table}

We evaluate the performance of the proposed DM-SBL algorithm using approximated noise-perturbed likelihood scores of DMPS and $\Pi$GDM, 
namely DM-SBL (DMPS) and DM-SBL ($\Pi$GDM), respectively. 
The analysis focuses on the estimated channel NMSE under various setting of SNR, SIR,  
number of channel taps and channel virtual length. 

In the assumed scenario, the carrier frequency of the LFM signal is the same as that of the communication signal, making it impossible to eliminate the interference through frequency-domain filtering. The structured interference $\mathbf{n}$ is randomly extracted from a unit-amplitude LFM signal with duration of 2 seconds, bandwidth of 1 kHz, and the initialized starting phase is randomly selected. To generate the channel, we adopt an underwater acoustic channel setting similar to that of 
\cite{wan2022joint, berger2009sparse}, there are $p_0$ paths, the arrival time between two adjacent paths 
follows an exponential distribution with mean value 3 ms and the amplitude of each path follows Rayleigh 
distribution with mean power decreasing by 20 dB in 30 ms delay spread. 
The pilot sequence $\bar{\mathbf{x}}$ is a randomly generated sequence modulated by BPSK. In the whole simulation part, we choose the length of pilot to ensure that the number of measurements $M$ is 200.  
A summary of the parameters used in the experiments are provided in Table \ref{para}. The SNR and SIR are defined as  
\begin{align}
{\rm SNR} = 10 \log_{10} \frac{\| \mathbf{A}\mathbf{h} \Vert^2_2 }{\|\boldsymbol{\epsilon}\Vert^2_2},{\rm SIR} = 10 \log_{10} \frac{\| \mathbf{A}\mathbf{h} \Vert^2_2 }{\|\mathbf{n}\Vert^2_2},
\end{align}
respectively.

To make performance comparision, the following benchmark algorithms are implemented:
\begin{itemize}
  \item \textbf{MMSE}: Basic channel estimator based on minimum mean squared error (MMSE) criterion 
  by assuming that $\mathbf{n} + \boldsymbol{\epsilon}$ is AWGN. 
  \item \textbf{EM-BGGAMP} \cite{6556987}: 
  The noise and interference are assumed to be AWGN, and the prior of channel $\mathbf{h}$ is modeled by 
  Bernoulli-Gaussian distribution. 
  We use the publicly available Matlab implementation \cite{amp} with adaptive damping and internal 
  EM steps. 
  \item \textbf{VAMP} \cite{8713501}: 
  The noise and interference are assumed to be AWGN, we use the publicly available Matlab implementation \cite{amp} with adaptive damping. 
  \item \textbf{OMP} \cite{cai2011orthogonal}: 
  Classic compressed sensing algorithm based on greedy method, where the
  sparsity of the channel $\mathbf{h}$ is known. 
  \item \textbf{SBL} \cite{tipping2001sparse}: 
  The noise and interference are assumed to be AWGN, the prior of the channel $\mathbf{h}$ is assumed to be 
  Gaussian distributed and the sparsity is controlled by its variance. 
\end{itemize}

\begin{figure*}
  \centering
  \subfloat[$t$ = 0.998]{\includegraphics[width=2in]{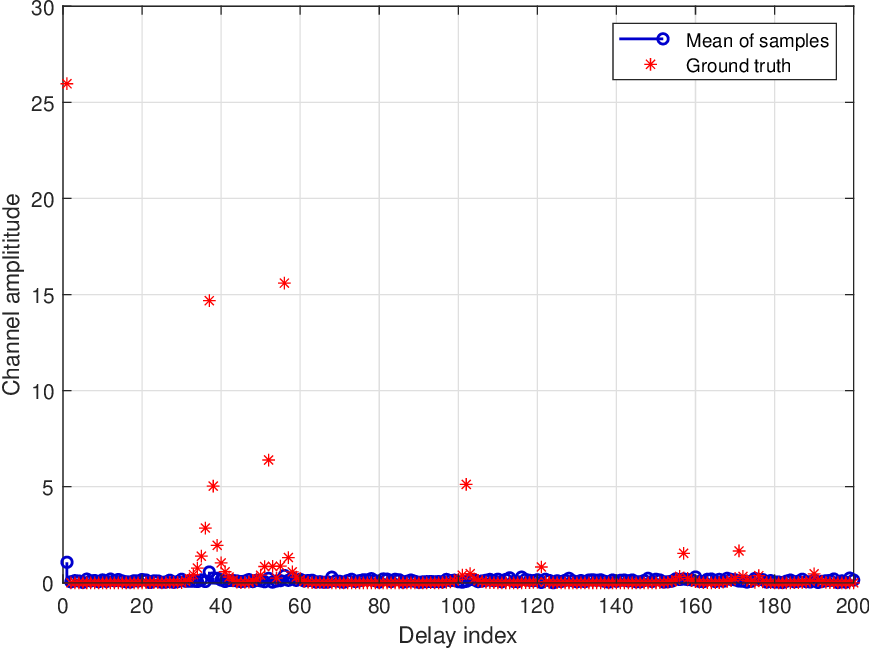}}\hspace{0.05mm}
  \subfloat[$t$ = 0.98]{\includegraphics[width=2in]{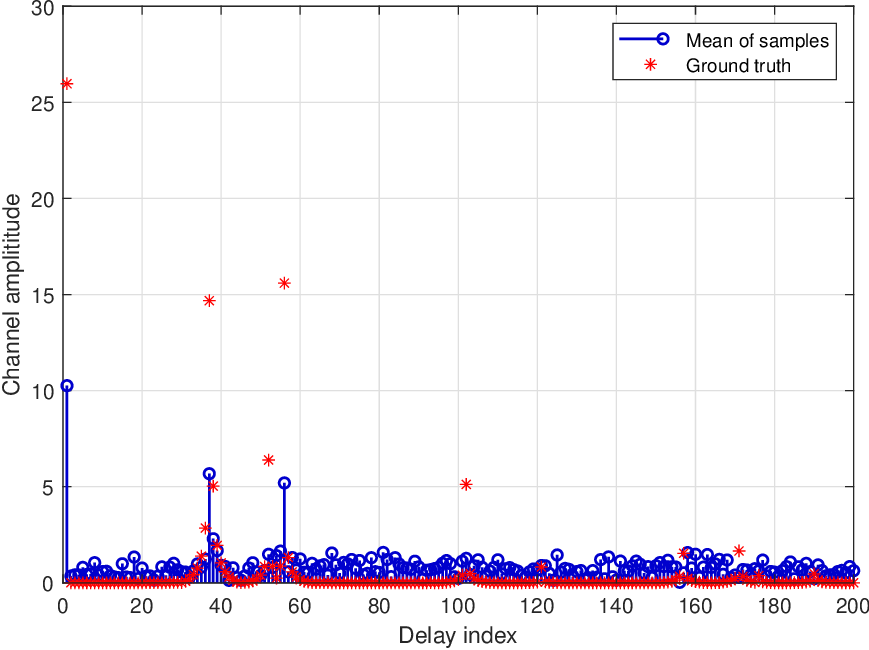}}\\
  \subfloat[$t$ = 0.8]{\includegraphics[width=2in]{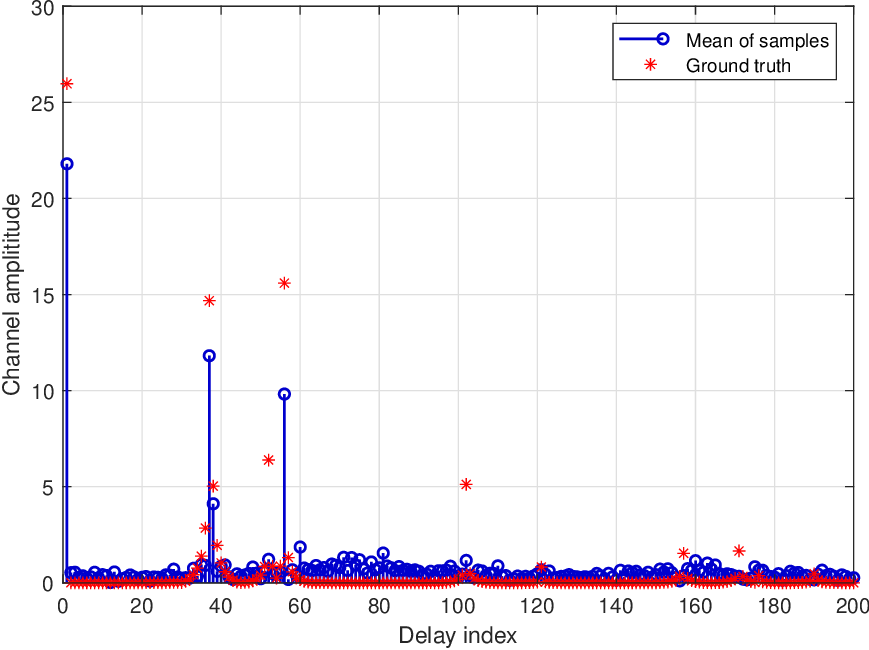}}\hspace{0.05mm}
  \subfloat[$t$ = 0]{\includegraphics[width=2in]{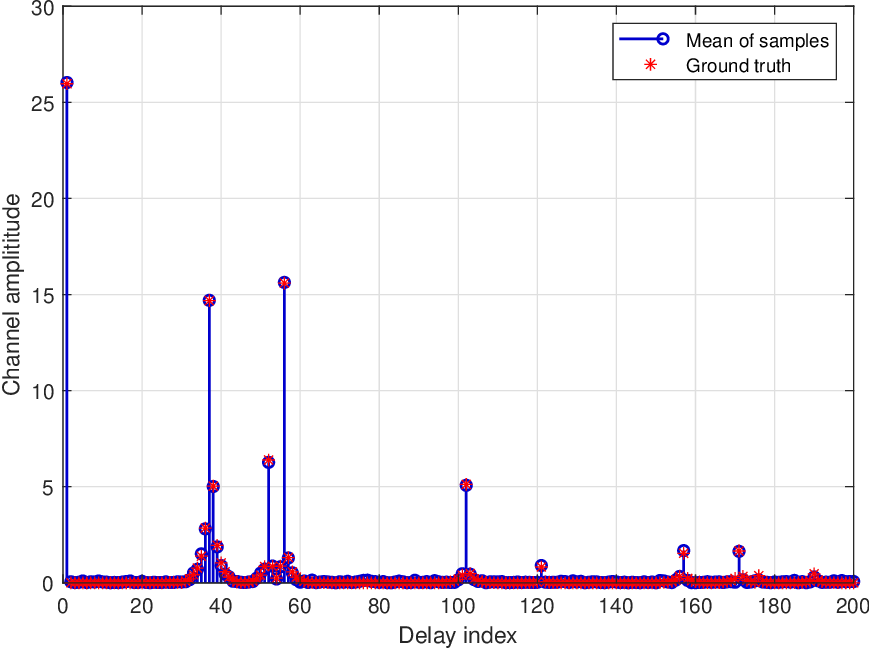}}
  \caption{Amplitude of $\sum_{i=1}^{K} \mathbf{h}_t^{(i)}/ (K \alpha(t))$ and ground truth in DM-SBL ($\Pi$GDM): (a) $t=0.998$; (b) $t=0.98$; (c) $t=0.8$; (d) $t=0$.}
  \label{sample_t}
\end{figure*}

\begin{figure*}[]
  \centering
  \subfloat[Ground truth]{\includegraphics[width=2.6in]{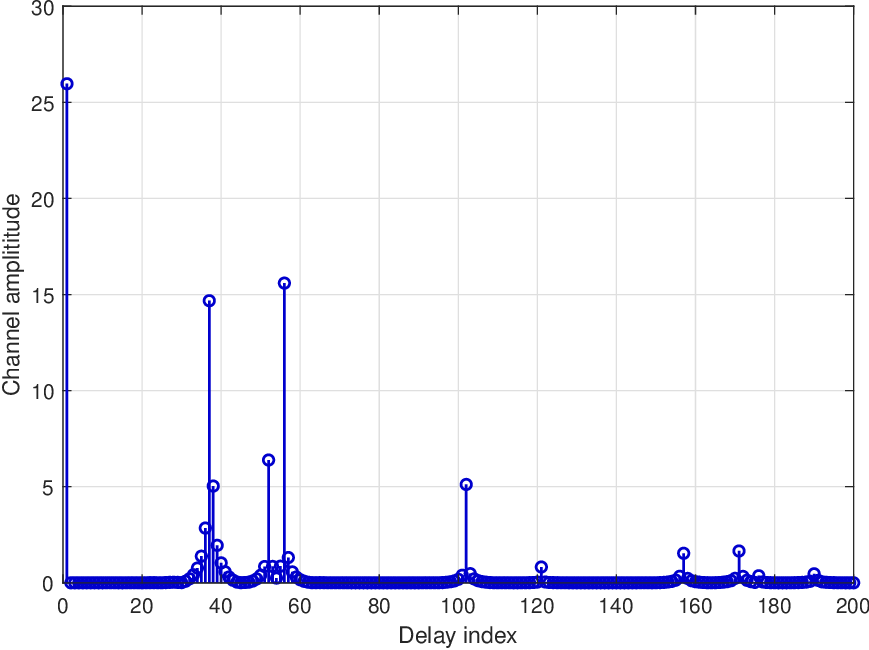}}
  \subfloat[DM-SBL (DMPS)]{\includegraphics[width=2.6in]{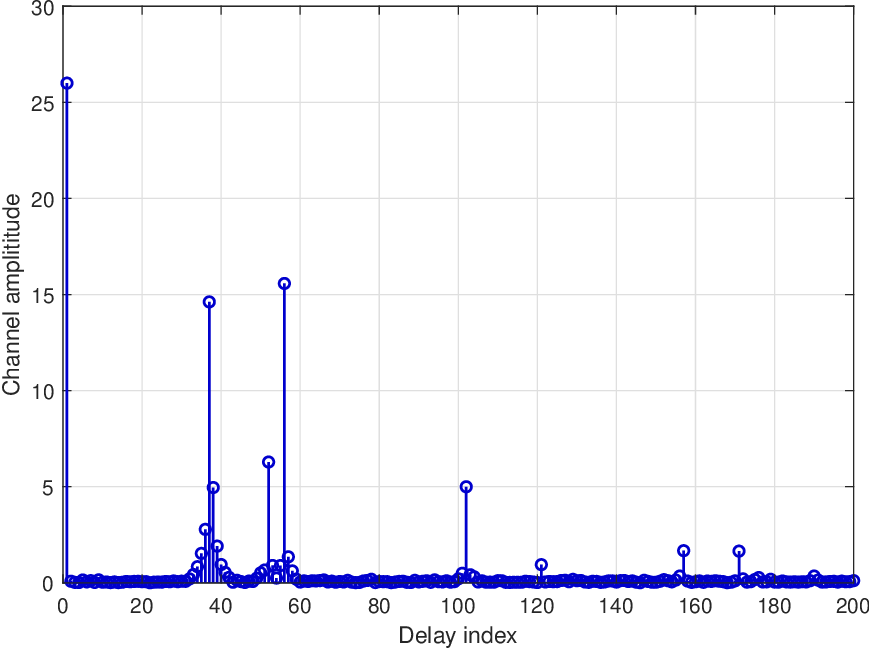}}\\
  \subfloat[DM-SBL ($\Pi$GDM)]{\includegraphics[width=2.6in]{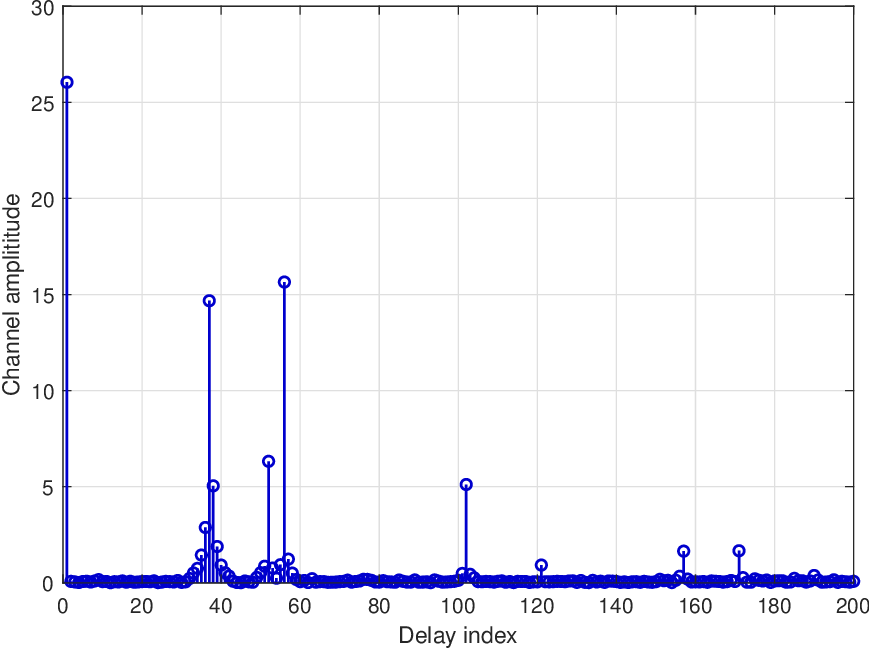}}
  \subfloat[SBL]{\includegraphics[width=2.6in]{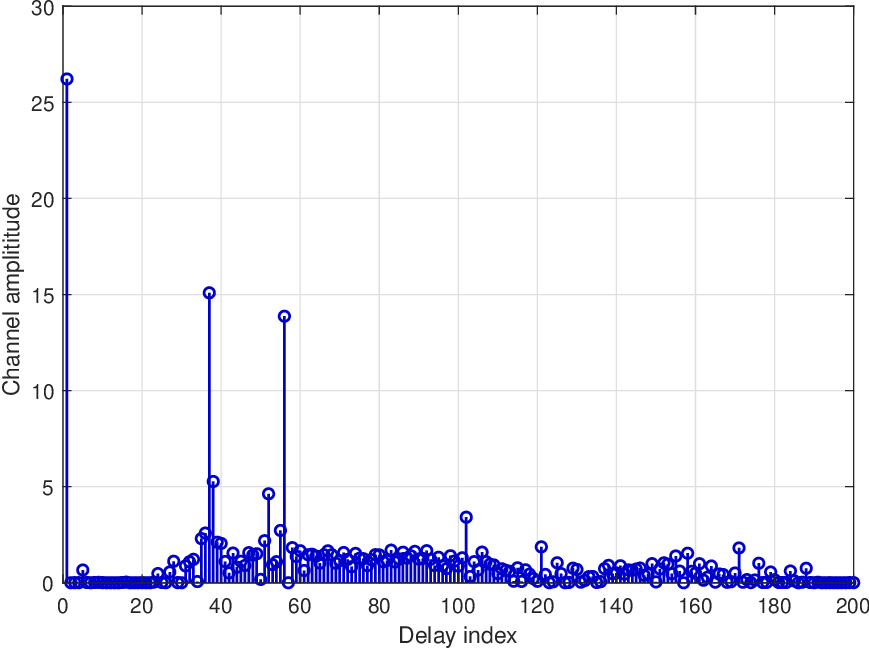}}\\
  \subfloat[EM-BGGAMP]{\includegraphics[width=2.6in]{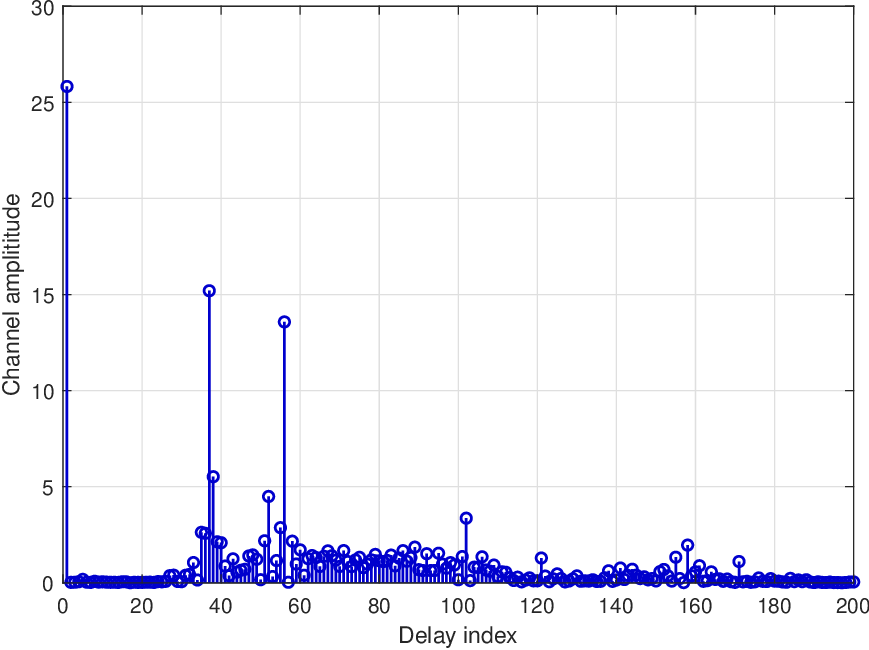}}
  \subfloat[VAMP]{\includegraphics[width=2.6in]{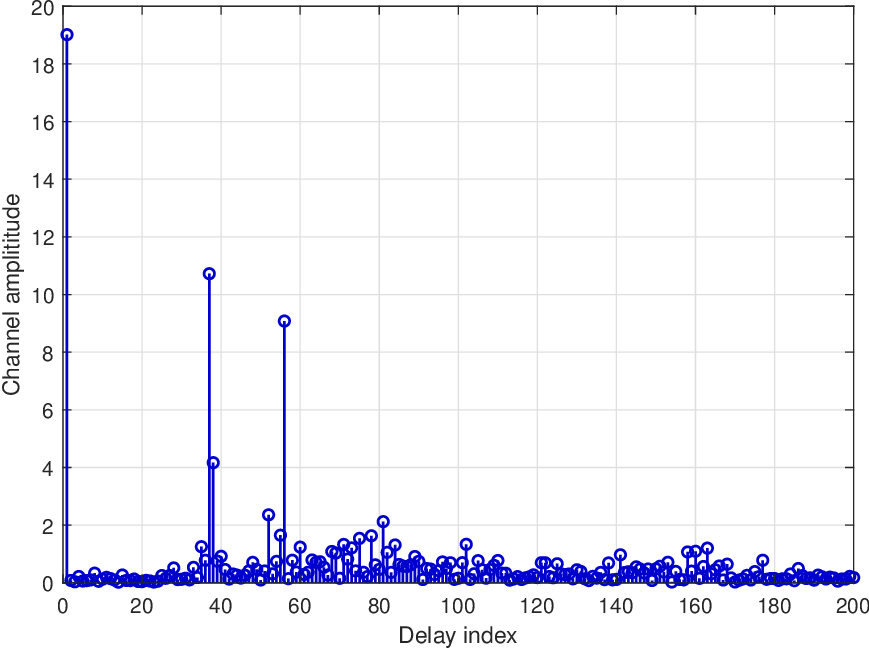}}\\
  \subfloat[OMP]{\includegraphics[width=2.6in]{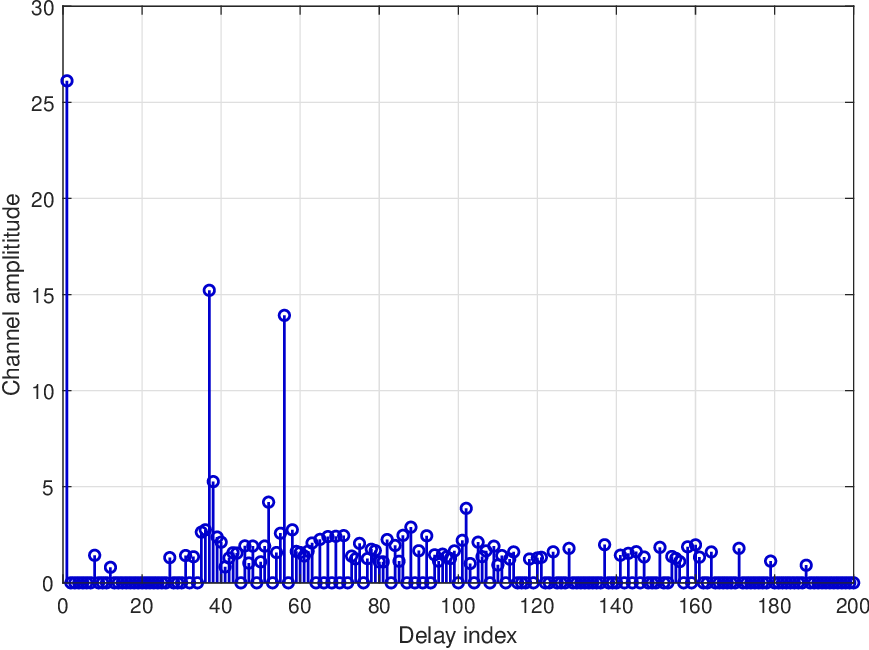}}
  \subfloat[MMSE]{\includegraphics[width=2.6in]{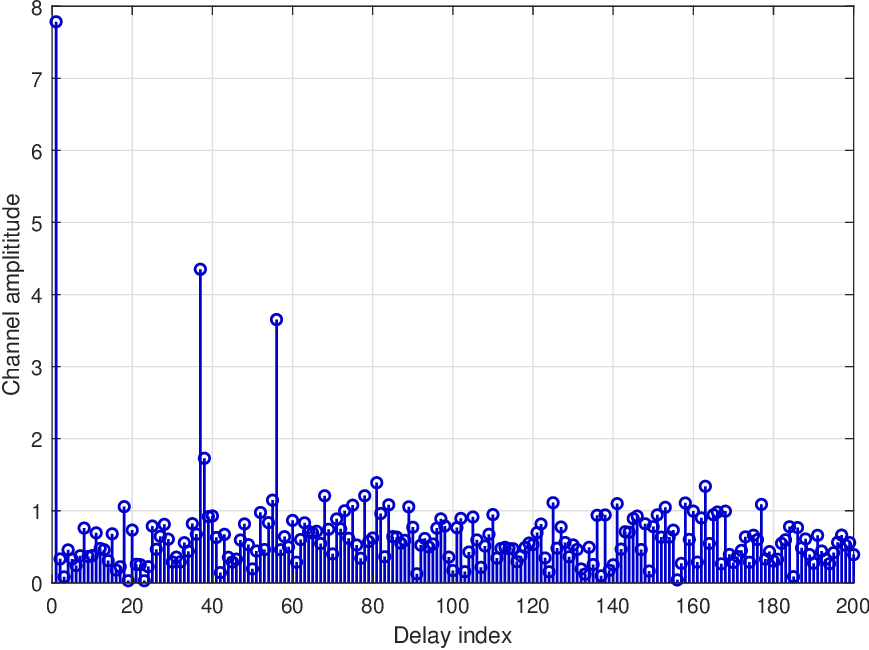}}
  \caption{
      Ground truth and estimated channel amplitude in a single realization, 
      $p_0=10$, $L=200$, SNR = 30 dB and SIR = 5 dB. 
      (a) Ground truth; (b) DM-SBL (DMPS); (c) DM-SBL ($\Pi$GDM); (d) SBL; 
      (e) EM-BGGAMP; (f) VAMP; (g) OMP; (h) MMSE.
  }
  \label{h_ex}
\end{figure*}
\subsection{Channel Estimation Results in a Single Realization}
First, we report the estimation results of all the algorithms in a single realization. The number of channel paths $p_0$ is 
10, the maximum virtual channel length $L$ is 200, SNR = 30 dB, SIR = 5 dB. The number of samples $K$ for DM-SBL is 256 and the number of sampling time step $T$ is 500. 
To better visualize the reverse diffusion process, taking DM-SBL (GDM) as an example, the magnitudes of the mean of samples at times $t = 0.998$, $t = 0.98$, $t = 0.8$ and $t=0$ over the course of the reverse process are shown in Fig. \ref{sample_t}. 
The paths with dominant energy appear first, while the paths with smaller energy gradually emerge later. 
The amplitude of ground truth of the channel is shown in Fig. \ref{h_ex} (a), estimated channel obtained by DM-SBL (DMPS), DM-SBL ($\Pi$GDM), SBL, EM-BGGAMP, VAMP, 
OMP and MMSE are shown in Fig. \ref{h_ex} (b)-(h), respectively. 
The NMSEs of all the algorithms are -29.95 dB, -30.47 dB, -8.90 dB, 
-9.76 dB, -7.73 dB, -7.18 dB, -2.32 dB, respectively. 
It can be seen that the estimated channel amplitudes obtained by both DM-SBL (DMPS) and DM-SBL ($\Pi$GDM) match the ground truth well, while other methods estimate spurious taps whose channel tap amplitudes are in fact zero. 
Note that in this setup, because the observations are heavily contaminated by non-Gaussian interference,  traditional methods which ignore the interference are not competitive with the DM-SBL methods, which makes sense
as DM-SBL exploits the structure of the interference.

\begin{figure*}[]
  \centering
  \subfloat[]{\includegraphics[width=2in]{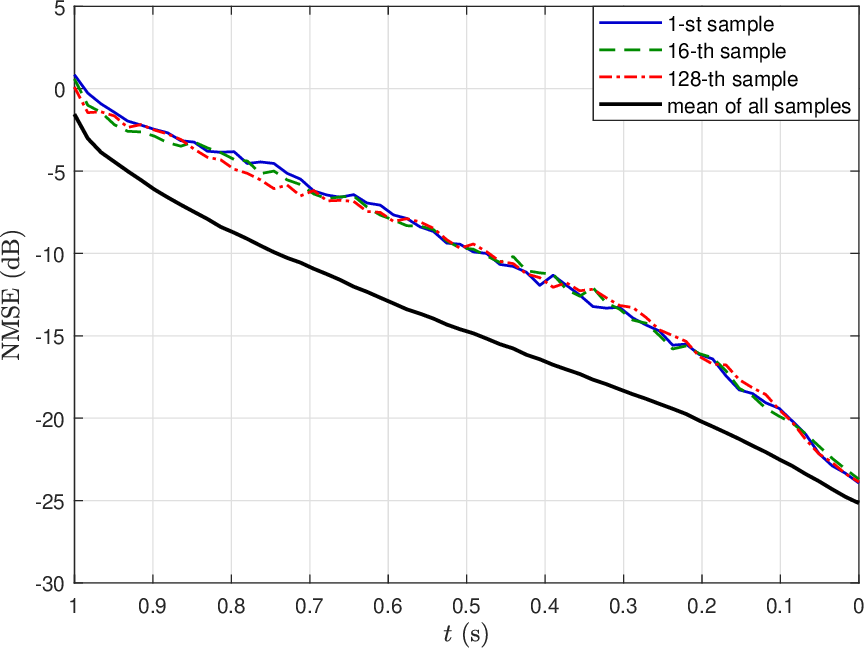}}\hspace{0.2mm}
  \subfloat[]{\includegraphics[width=2in]{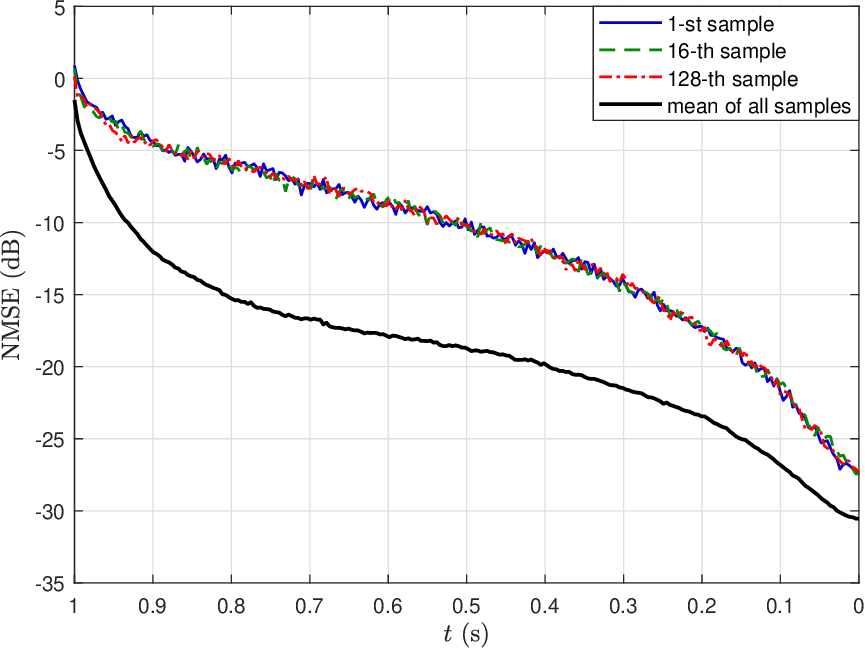}}\hspace{0.2mm}
  \subfloat[]{\includegraphics[width=2in]{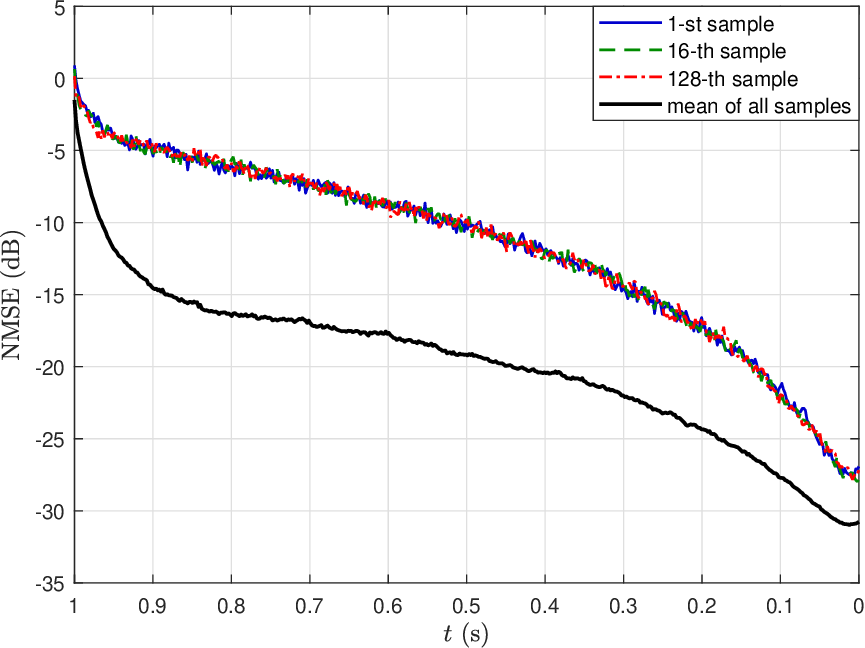}}\\
  \subfloat[]{\includegraphics[width=2in]{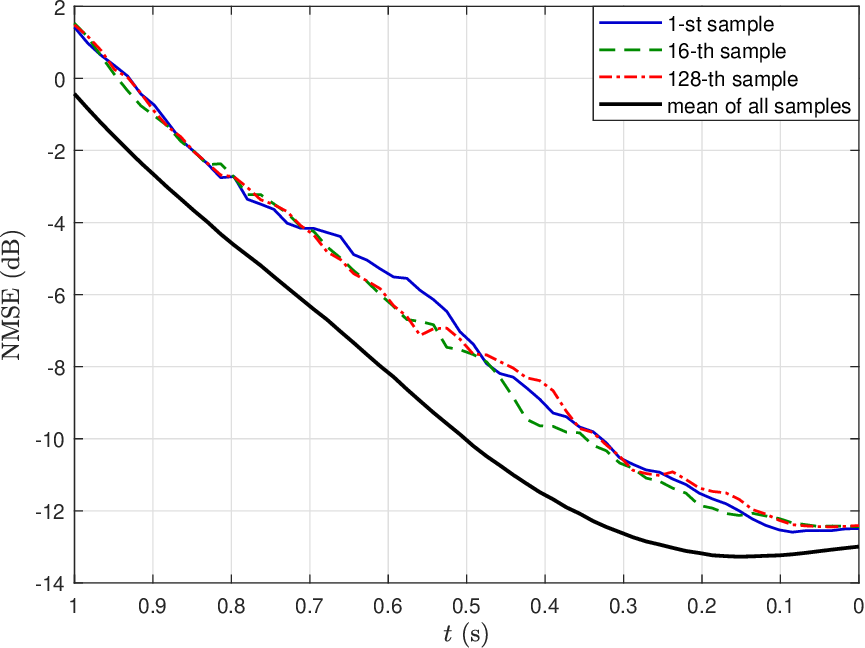}}\hspace{0.2mm}
  \subfloat[]{\includegraphics[width=2in]{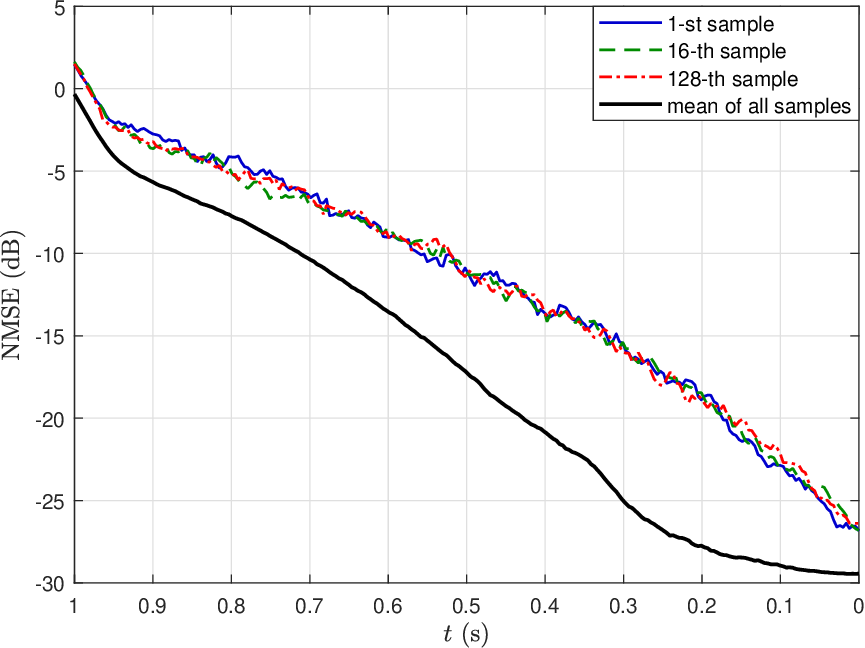}}\hspace{0.2mm}
  \subfloat[]{\includegraphics[width=2in]{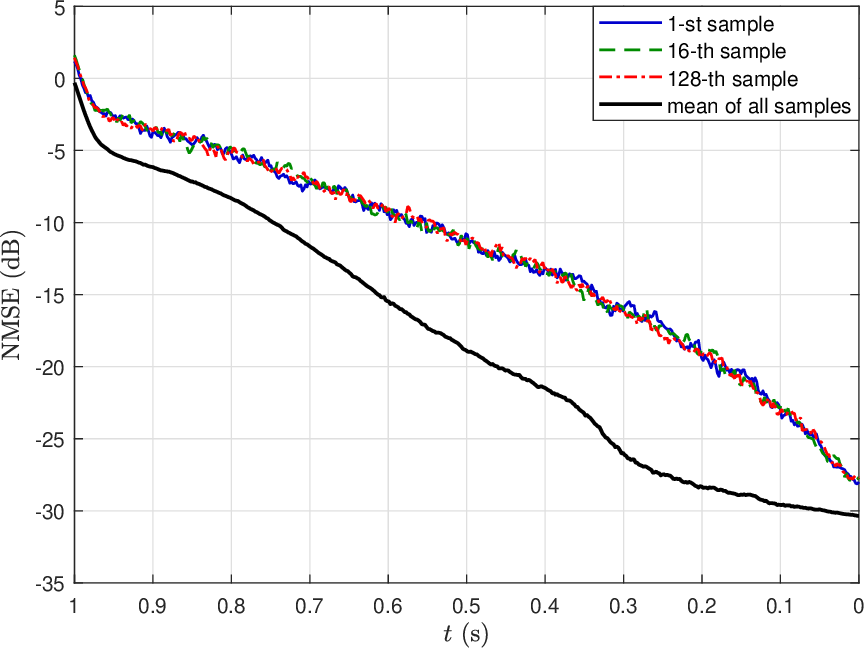}}
  \caption{
    Sampled channel NMSE versus time $t$ for different setting of sampling steps $T$ using 
    DM-SBL (DMPS) and DM-SBL ($\Pi$GDM), 
    number of samples $K$ = 256 in each time step, $p_0$ = 10, $L$ = 200, SNR = 30 dB and SIR = 5 dB.  
    (a) DM-SBL (DMPS), $T=60$; (b) DM-SBL (DMPS), $T=250$; (c) DM-SBL (DMPS), $T=500$; 
    (a) DM-SBL ($\Pi$GDM), $T=60$; (b) DM-SBL ($\Pi$GDM), $T=250$; (c) DM-SBL ($\Pi$GDM), $T=500$. 
  }
  \label{sample_nmse}
\end{figure*}
\subsection{Sampling Steps $T$ on the Channel Estimation}

Next, we investigate how the sampling steps $T$ affect the channel estimation 
performance and how individual samples of $\mathbf{h}$ and the 
mean of the samples converge to the ground truth. The NMSE of $\mathbf{h}_t^{(i)}/\alpha(t)$ for specific $i=1,16,128$ and 
$\sum_{i=1}^{K} \mathbf{h}_t^{(i)}/ (K \alpha(t))$ versus $t$ for different settings of $T$ using DM-SBL (DMPS) and 
DM-SBL ($\Pi$GDM) are shown in Fig. \ref{sample_nmse}. 
It can be observed that although different samples are initially generated using a zero mean Gaussian distribution, 
$\mathbf{h}_t^{(i)}/\alpha(t)$ converges to the ground truth at an almost same speed due to the 
correlation induced by the measurement model. 
The mean of all samples always leads to lower NMSE compared with individual samples, demonstrating 
the importance of using multiple simultaneously sampling process. 
With $T$ increasing from 60 to 250, the NMSE of $\mathbf{h}_0$ slightly decreases for 
DM-SBL (DMPS) and largely decreases for DM-SBL ($\Pi$GDM). When $T$ increases to 500, the 
NMSEs of both DM-SBL (DMPS) and DM-SBL ($\Pi$GDM) show little difference with that of $T$ = 250. 
Increasing the number of sampling steps actually reduces the discretization error introduced by numerical methods for solving the reverse SDE and makes the step-by-step denoising process more accurate. Further increasing $T$ does not lead to a significant improvement in performance here. However, in Monte Carlo trials, we have found that increasing $T$ enhances the stability of the algorithm.

\begin{figure*}[htpb]
  \centering
  \subfloat[]{\includegraphics[width=2.8in]{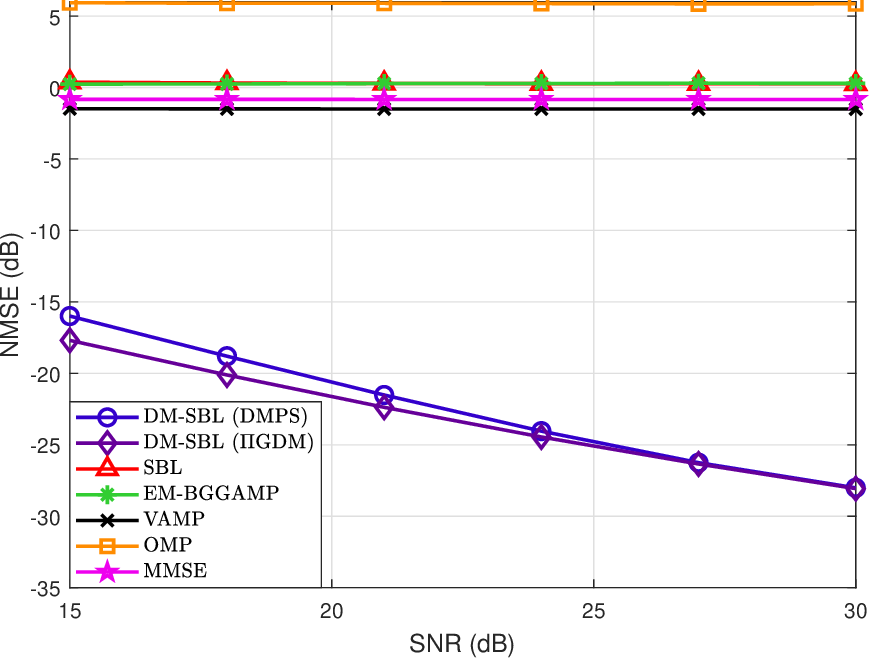}}\hspace{0.2mm}
  \subfloat[]{\includegraphics[width=2.8in]{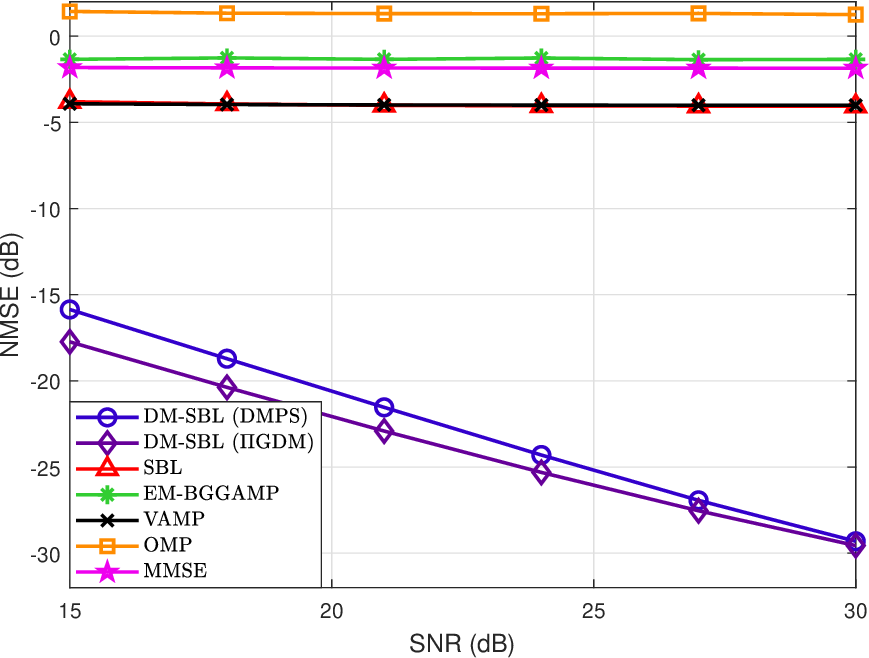}}\\
  \subfloat[]{\includegraphics[width=2.8in]{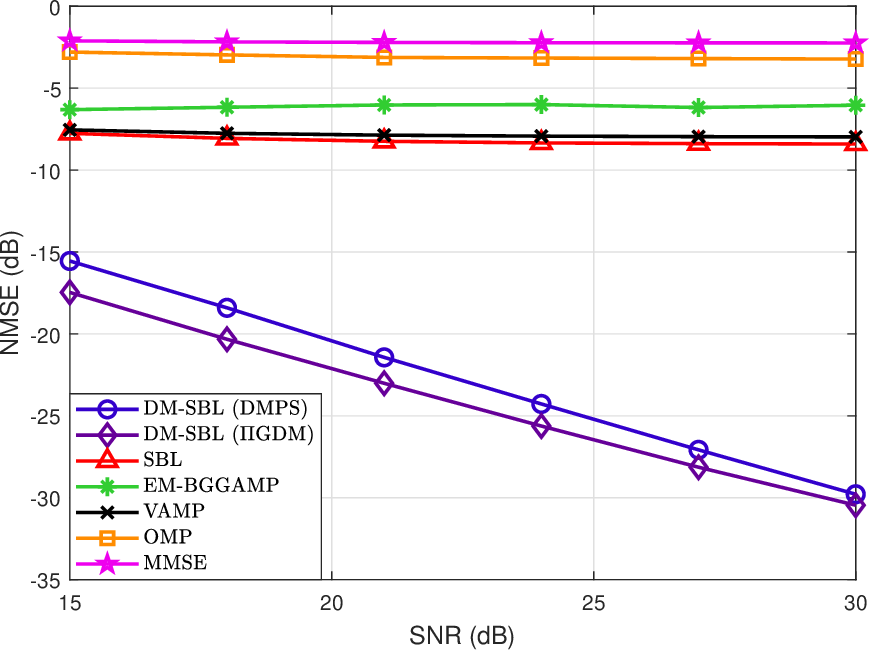}}\hspace{0.2mm}
  \subfloat[]{\includegraphics[width=2.8in]{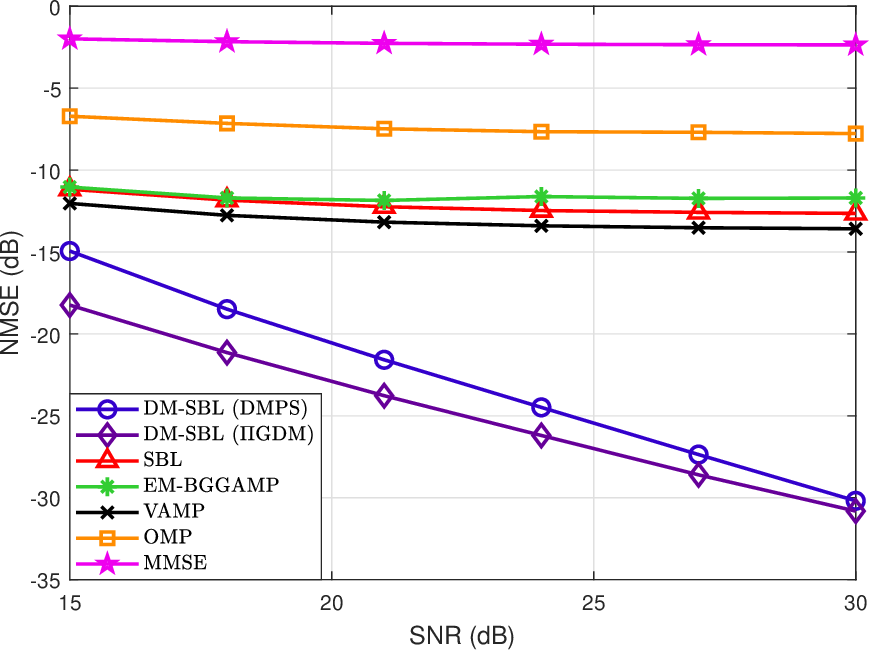}}
  \caption{
    The NMSEs of the channel estimates versus SNR for all algorithms, where $L = 200$, $p_0 = 10$, 
    $\mu$ = 1 for both DM-SBL (DMPS) and DM-SBL ($\Pi$GDM), 
    $\kappa$ = 0.18 for DM-SBL (DMPS) and 0.05 for DM-SBL ($\Pi$GDM), SIR = : (a) -5 dB; (b) 0 dB; (c) 5 dB; (d) 10 dB.
  }
  \label{L200}
\end{figure*}
\subsection{Channel Estimation Results versus SNR}
The channel NMSE versus SNR at SIR = -5, 0, 5, 10 dB for $p_0$ = 10 and $L = 200$ are shown in Fig. \ref{L200} (a)-(d), respectively. 
As the SIR decreases, the channel estimation performance of the proposed two methods remain stable across different SNRs, while the performances of other algorithms deteriorates rapidly. 
At an SIR of -5 dB, all bench algorithms fail to obtain any accurate information about the channel, with their NMSEs approaching or even exceeding 0 dB. In contrast, DM-SBL is still able to provide relatively accurate channel estimation.
For a given SIR, increasing SNR does not provide significant gains for other algorithms. This is because the SIR is much lower than the SNR, and interference is the major factor that affects the performance. The NMSE of DM-SBL (DMPS) is close to that of DM-SBL (GDM), with the latter performing slightly better than the former. Other benchmark algorithms, such as EM-BGGAMP, SBL, VAMP, can barely work at SIR = 10 dB. 

\begin{figure*}[htpb]
  \centering
  \subfloat[]{\includegraphics[width=2.8in]{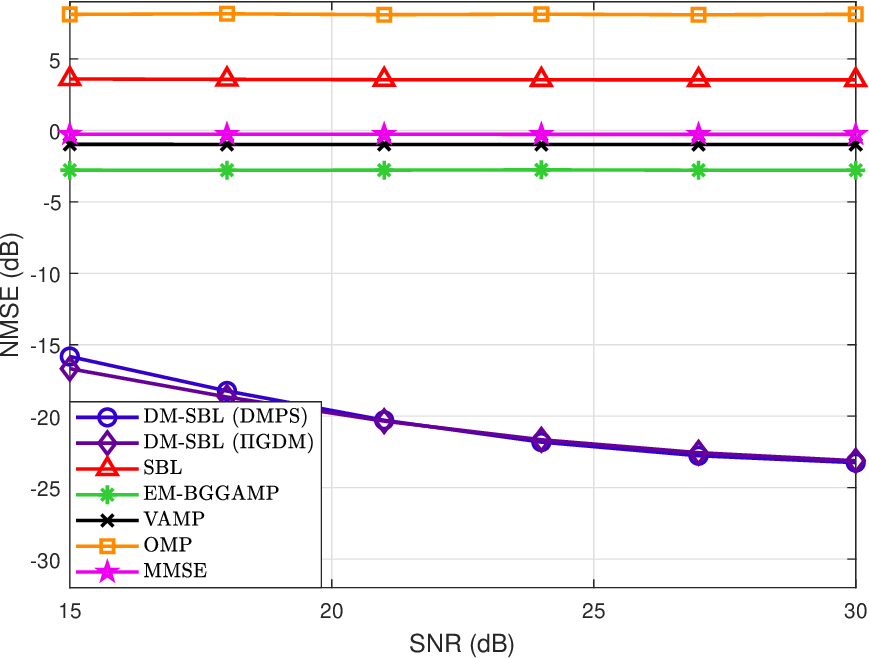}}\hspace{0.2mm}
  \subfloat[]{\includegraphics[width=2.8in]{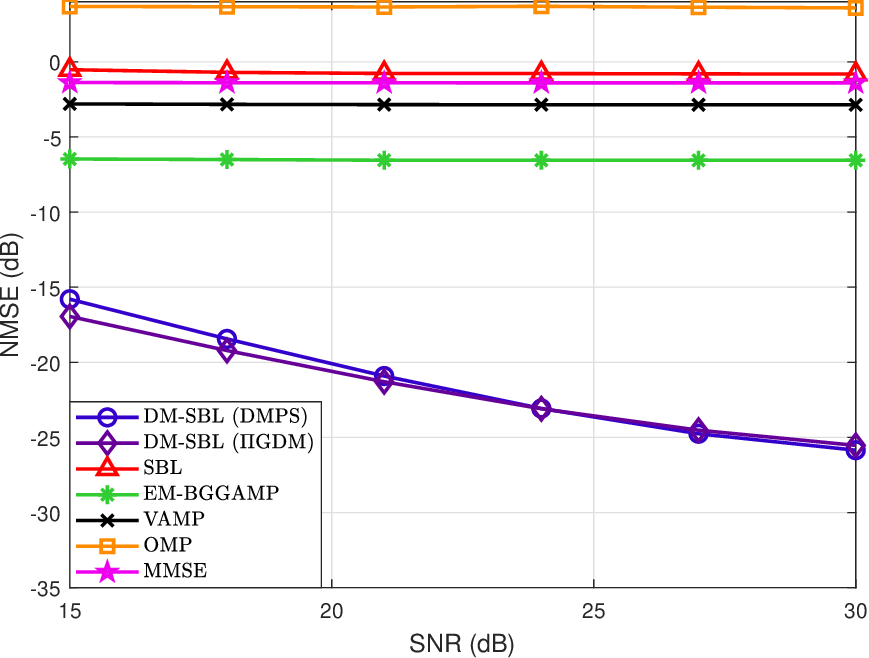}}\\
  \subfloat[]{\includegraphics[width=2.8in]{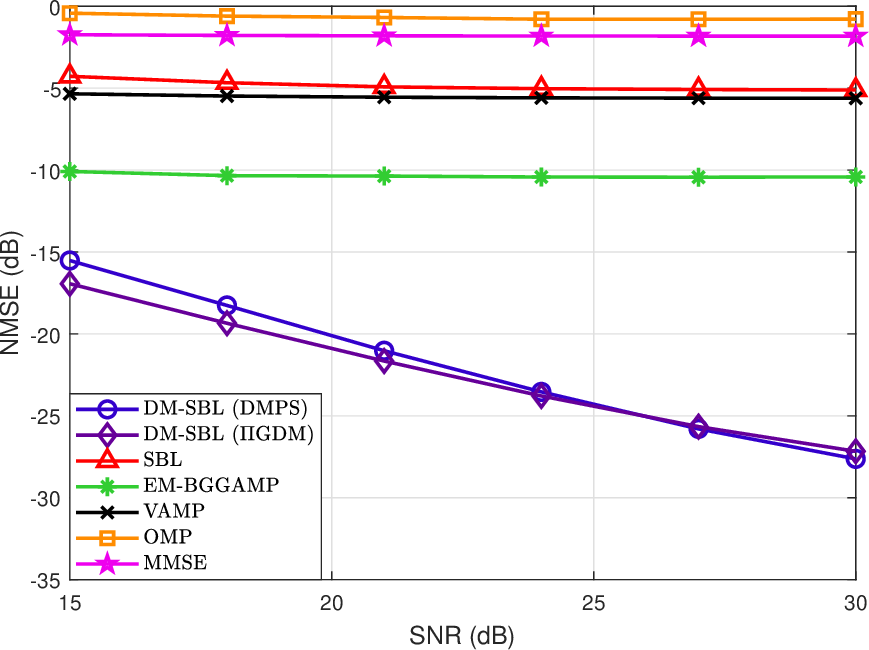}}\hspace{0.2mm}
  \subfloat[]{\includegraphics[width=2.8in]{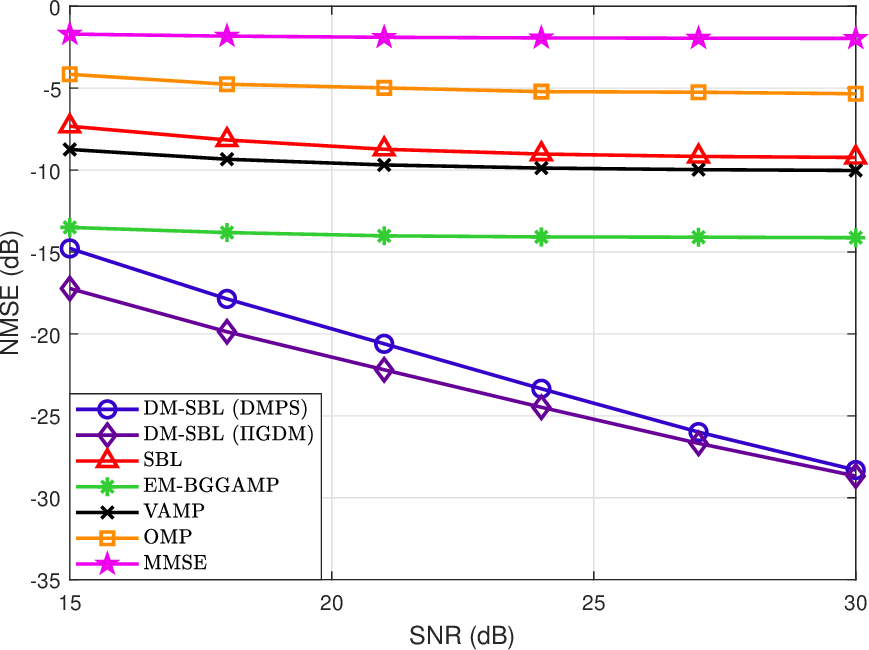}}
  \caption{
    The NMSEs of the channel estimates versus SNR for all algorithms, where $L = 300$, $p_0 = 15$, 
    $\mu$ = 1 for both DM-SBL (DMPS) and DM-SBL ($\Pi$GDM), 
    $\kappa$ = 0.15 for DM-SBL (DMPS) and 0.05 for DM-SBL ($\Pi$GDM), SIR = : (a) -5 dB; (b) 0 dB; (c) 5 dB; (d) 10 dB.
  }
  \label{L300}
\end{figure*}

Fig. \ref{L300} (a)-(d) show the channel NMSE versus SNR at SIR = -5, 0, 5, 10 dB, where all parameters are fixed except that the number of path $p_0$ and the number of virtual paths $L$ are increased to $p_0$ = 15 and $L = 300$. Compared with Fig. \ref{L200}, both DM-SBL (DMPS) and DM-SBL ($\Pi$GDM) show some performance degradation due to the 
increase of virtual channel length $L$ and number of channel paths $p_0$. Under different SNR and SIR settings, the performance of DM-SBL ($\Pi$GDM) is slightly better than that of DM-SBL (DMPS). Among the other comparison algorithms, only EM-BGGAMP performs somewhat well when the SIR is 5 dB or 10 dB.

Regarding the complexity of the proposed methods, with $L$ = 200, $M$ = 200, $T$ = 500, and $K$ = 256, the average computation times of DM-SBL (DMPS) and DM-SBL ($\Pi$GDM) are 11 seconds and 18 seconds, respectively. For $L$ = 300, the corresponding average computation times are 12 seconds and 18 seconds. These results are obtained on a machine equipped with an NVIDIA RTX 4090 GPU and an Intel i9-9900K CPU. Compared to DM-SBL (DMPS), DM-SBL ($\Pi$GDM) requires slightly more computation time due to the need for Jacobian matrix calculations. However, DM-SBL ($\Pi$GDM) also performs slightly better than DM-SBL (DMPS).

\section{Conclusion}\label{conclu}
In this paper, we proposed a versatile framework for channel estimation in the presence of structured interference, combining score-based diffusion models with the SBL. The score of the structured interference is learned through a neural network, while the channel score is derived analytically. Leveraging the perturbed likelihood approximations used in DMPS and $\Pi$GDM, we developed two algorithms: DM-SBL (DMPS) and DM-SBL ($\Pi$GDM). 
Numerical results show that their performance is significantly better than the traditional methods which ignores the interference, especially when the interference is very strong. 

Current score-based conditional sampling channel estimation methods require the neural network to learn the score for the channel. The proposed SM-SBL framework allows seamlessly to adapt to various channel types without the need for additional training of channel scores. This flexibility enables rapid deployment across diverse channel environments and avoids constraints imposed by specific channel distributions. 

Future research should aim to enhance the computational efficiency of the algorithms, investigate more advanced network architectures to handle increasingly complex interference scenarios, address symbol demodulation under structured interference. 

\begin{appendices}
\section{Details of the score neural network}\label{NNAppendix}

\begin{figure*}[htbp]
  \centering
  \includegraphics[width=6.5in]{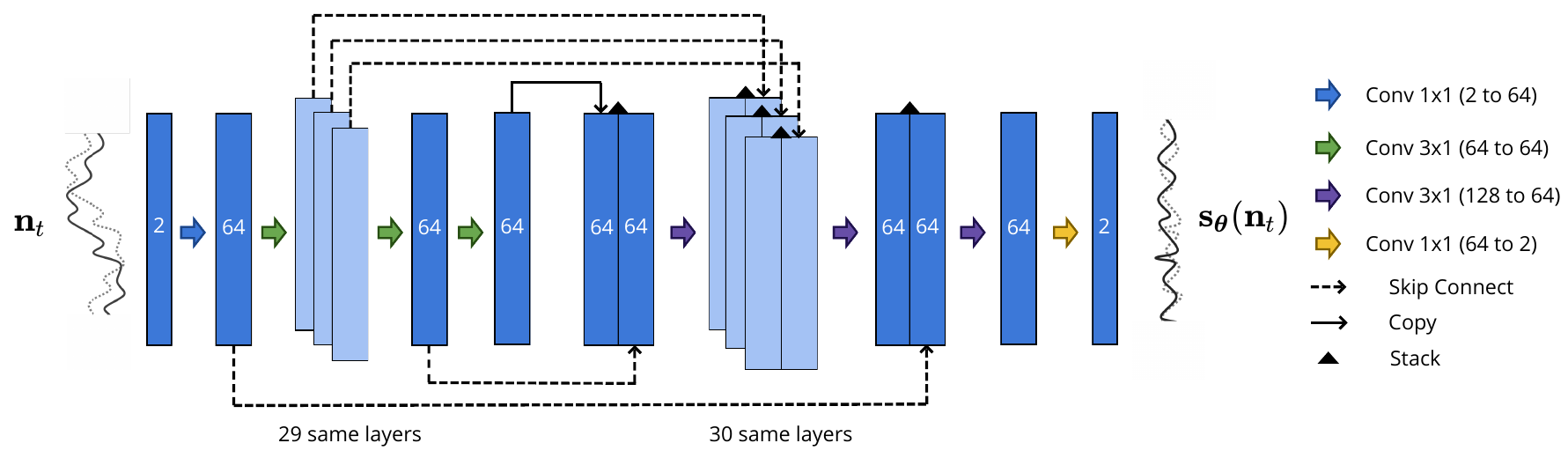}
  \caption{
        Structure of the score estimating network $\mathbf{s}_{\boldsymbol{\theta}}$ using a full conditional network with a similar framework to U-net. 
  }
  \label{score}
\end{figure*}

The score neural network  shown in Fig. \ref{score}, adopts a structure similar to U-Net \cite{ronneberger2015u}, featuring fully convolutional layers. The input, represented as two stacked channels for the real and imaginary parts is projected into 64 feature dimensions through a convolutional layer. Both the encoding and decoding process consist of 32 identical blocks of convolutional layer, ReLU activation and skip connection to stack features with that in decoding process, which enhances gradient flow and preserve feature integrity. The final convolutional layer projects the output into a two-dimensional space, representing the estimated score.

\end{appendices}



\small
\bibliography{IEEEabrv,myrefs}
\bibliographystyle{IEEEtran}

\end{document}